\documentclass[preprint,5p,authoryear]{elsarticle}

\usepackage{pgfplots}
\usepackage{amsmath}
\usepackage{subfigure}
\usepackage{placeins} 
\usepackage{ragged2e} 
\usepackage{multirow}
\usepackage{natbib}

\newcommand{\sm}[1]{\rm{{\scriptsize #1}}}

\newcommand{\sun}{{\odot}}

\newcommand{\Msol}{\rm{M}_{\sun}}

\newcommand{\yr}{\rm{yr}}

\usetikzlibrary{shapes.geometric, arrows}
\tikzstyle{startstop} = [rectangle, rounded corners, minimum width=3cm, minimum height=1cm,text centered, draw=black]
\tikzstyle{io} = [trapezium, trapezium left angle=70, trapezium right angle=110, minimum width=3cm, minimum height=1cm, text centered, draw=black, fill=blue!30]
\tikzstyle{process} = [rectangle, minimum width=3cm, minimum height=1cm, text centered, draw=black, fill=green!30]
\tikzstyle{flash} = [ fill=blue,text=white,minimum width=9cm,minimum height=1cm, anchor=north west,rotate=90]
\tikzstyle{arrow} = [thick,->,>=stealth]
\tikzstyle{line} = [thick,--,>=stealth]

\definecolor{rosso}{RGB}{220,57,18}
\definecolor{giallo}{RGB}{255,153,0}
\definecolor{blu}{RGB}{102,140,217}
\definecolor{verde}{RGB}{16,150,24}
\definecolor{viola}{RGB}{153,0,153}

\tikzstyle{chart}=[
    legend label/.style={font={\scriptsize},anchor=west,align=left},
    legend box/.style={rectangle, draw, minimum size=5pt},
    axis/.style={black,semithick,->},
    axis label/.style={anchor=east,font={\tiny}},
]

\tikzstyle{bar chart}=[
    chart,
    bar width/.code={
        \pgfmathparse{##1/2}
        \global\let\bar@w\pgfmathresult
    },
    bar/.style={very thick, draw=white},
    bar label/.style={font={\bf\small},anchor=north},
    bar value/.style={font={\footnotesize}},
    bar width=.75,
]

\tikzstyle{pie chart}=[
    chart,
    slice/.style={line cap=round, line join=round, very thick,draw=white},
    pie title/.style={font={\bf}},
    slice type/.style 2 args={
        ##1/.style={fill=##2},
        values of ##1/.style={}
    }
]

\pgfdeclarelayer{background}
\pgfdeclarelayer{foreground}
\pgfsetlayers{background,main,foreground}

\usepackage{tablefootnote}
\usepackage{float}
\pgfplotsset{compat=1.3}

\begin{document}

\begin{frontmatter}

\title{A GPU accelerated Barnes-Hut Tree Code for FLASH4}

\author{Gunther Lukat\corref{cor1}}
\ead{glukat@hs.uni-hamburg.de}
\author{Robi Banerjee}
\ead{banerjee@hs.uni-hamburg.de}
\cortext[cor1]{Corresponding author}
\address{Hamburger Sternwarte, Universit\"at Hamburg, Gojenbergsweg 112, 21029 Hamburg}

\begin{abstract}
We present a GPU accelerated CUDA-C implementation of the Barnes Hut (BH)
tree code for calculating the gravitational potential on octree
adaptive meshes. The tree code algorithm is implemented within the
FLASH4 adaptive mesh refinement (AMR) code framework and therefore fully MPI parallel. We describe
the algorithm and present test results that demonstrate its accuracy
and performance in comparison to the algorithms available in the
current FLASH4 version. We use a MacLaurin spheroid to test the
accuracy of our new implementation and use spherical, collapsing cloud cores with effective AMR
to carry out performance tests also in comparison with previous
gravity solvers. Depending on the setup and the GPU/CPU ratio, we
find a speedup for the gravity unit of at least a factor of 3 and up to 60 in comparison
to the gravity solvers implemented in the FLASH4 code. 
We find an overall speedup factor for full simulations of at least factor 1.6 up to a factor of 10. 

\end{abstract}

\begin{keyword}
gravitation \sep hydrodynamics \sep methods: numerical \sep stars: formation

\end{keyword}

\end{frontmatter}

\section{Introduction}
\label{introduction}
Self-Gravity is a key phenomena in many Astrophysical Simulations, hence the solution of Poisson's equation of the gravitational potential $\phi(x)$ for a given density distribution $\rho(x)$
\begin{equation} 
\nabla^2 \phi(x)=4\pi G \rho(x)
\end{equation}
is one of the main functions in these simulations. The direct method for evaluating the gravitational potential at position $x_i$ of a point mass $m_i$ requires the evaluation of all pairwise interactions in the system. 
\begin{equation}
\Phi(x_{\mathrm{i}})=- \frac{1}{2} G \sum_{\substack{\mathrm{i}=1\\
    \mathrm{i}\not=\mathrm{j} }}^N
\frac{m_\mathrm{j}}{|\vec{r}_{\mathrm{i}} - \vec{r}_{\mathrm{j}}|} \, ,
\end{equation}
where, $m_\mathrm{j}$ is the mass of the jth particle and
$\vec{r}_\mathrm{i}$ and $\vec{r}_\mathrm{j}$ are the position vectors of particle i and j. While the direct method with its $O(N^2)$ arithmetic complexity is conceptually simple, it is obviously unsuitable for larger systems. Luckily several heuristic algorithms are available which require fewer operations within acceptable error bounds.\\ 
In general, one can distinguish between the grid-based algorithms and the tree-based algorithms for calculating the gravitational potential. We use the FLASH code package \citep{2000ApJS..131..273F, Fisher:2008bn}‚ and three of the implemented algorithms: the multipole solver, the multigrid solver and the tree solver.\\
The \emph{Multipole Poisson solver} is based on a multipolar expansion of the mass distribution around a certain center of expansion. Both accuracy and runtime can be controlled via the multipole cutoff value $l_{max}$. The multipole approach is by nature appropriate for systems with spherical mass distributions, such that a spherical harmonic expansion can be expected to reach high accuracy after a small number of terms. \citep{Couch:2013tra}\\
The \emph{Multigrid Poisson solver} is a modified version of the direct multigrid algorithm of Huang \& Greengard adapted to the FLASH4 grid structure \citep{2008ApJS..176..293R}. The \emph{Multigrid Poisson solver} is appropriate for general mass distributions.\\
The \emph{Tree Poisson solver} is based on the Barnes \& Hut tree code
where the implemented octree is an extension of the AMR mesh tree down
to the individual cells (based on the FLASH4.2.2 release). The tree Poisson solver is appropriate for general mass distributions. In the following sections we refer to the  Barnes \& Hut tree Poisson solver as CPU-BH tree solver.\\
Although the heuristic solvers reduce the computation time for the gravity module, they still require a large amount of computation time, often much more than the integration of the MHD equations.\\
In this paper, we developed a GPU accelerated Barnes \& Hut tree Poisson solver for astrophysical applications based on the CUDA runtime library \citep{NVIDIA:tc, Nickolls:2008vi}. Our GPU accelerated Barnes \& Hut tree code implements tree walks as well as tree builds solely on the GPU-Device. By accelerating the gravity module with the GPU accelerated  Barnes \& Hut tree code,  we measured a performance improvement of more than factor 20. In the following sections we refer to our GPU accelerated Barnes \& Hut tree Poisson solver as GPU-BH tree solver.\\

\section{Algorithm}
\label{algorithm}
\begin{figure}\centering
\begin{tikzpicture}

\draw (4,0) -- (4, 4);
\draw (4,4) -- (8, 4);
\draw (4,0) -- (8, 0);
\draw (8,0) -- (8, 4);

\node (CENT) at (6,2) [circle,draw,shade,minimum size=0.15] {};
\node (COM) at (5.6,0.7) [circle,draw,shade,minimum size=0.15] {};
\node (PARTICLE) at (1.5,1.7) [circle,draw,fill,minimum size=0.15] {};

\node[right of = COM , yshift = -0.3cm,xshift=-05] {C.O.M};
\node at (2,3) [] {\large $d > l/\theta +\delta$};

\draw[<->, very thick] (4,3.5) to node[above] {$l$} (8,3.5);

\draw[<->, very thick] (PARTICLE) to node[above] {$d$} (COM);

\draw[<->, very thick] (CENT) to node[right] {$\delta$} (COM);

\end{tikzpicture}
\caption{The Geometry of the Barnes \& Hut opening criterion used in the GPU-BH tree code.}
\label{opening_angles}
\end{figure}
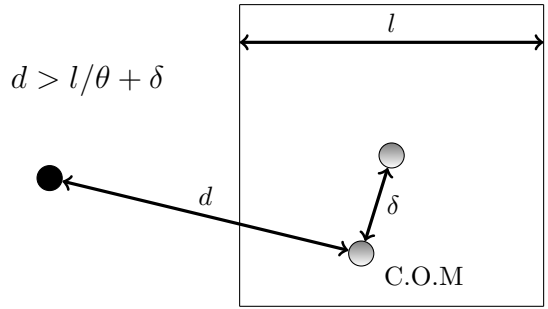

The  Barnes \& Hut tree algorithm for three-dimensions works by grouping ``particles'' \footnote{We use the term ``particles'' as a proxy for any mass-item, including grid cells.} using an octree structure \citep{Barnes:1986ed}. A single cube containing all particles surrounds the system.  This cube is then recursively divided into eight sub cells with each containing their own set of cells. The tree building method continues down in scale until only one particle is left for every sub cell. The tree construction can be done using a bottom-up i.e. inserting one particle at a time or a top-down approach by sorting the particles across divisions. Both methods take \emph{O(N log N)} time.\
For each tree node, the total mass and the center of mass is calculated. The force on a particle in the system is evaluated by ``walking'' down the tree. At each level, every node is tested against a test particle if it is distant enough for a force evaluation. If the node is too close, it is ``opened'' and the 8 children are selected for the same procedure. Various criteria exist to test whether a particle is sufficiently distant for a force evaluation. The most common and simplest criterion is based on the opening angle parameter $\theta$ \citep{Barnes:1986ed}.
If the size of a node is $l$ and the distance of the particle from the cell center of mass is $d$, the node can be accepted for a force evaluation if 
\begin{equation}\label{THETA1}
d > l/\theta.
\end{equation}
Smaller values of  $\theta$ lead to a higher accuracy cause by more node selected for opening. Typically a  $\theta$ value of  $1$ leads to errors around $1\%$ \citep{1987ApJS...64..715H}. In some cases, in which the center of mass is near the edge of a node, the basic criterion described in Eq.~\ref{THETA1} can produce large errors \citep{1994JCoPh.111..136S} Various alternatives are given in literature to avoid this problem. \citep{1994JCoPh.111..136S, Barnes:1994tf} Here, we adopt the opening angle parameter described by \cite{Barnes:1994tf} with
\begin{equation}\label{THETA2}
d > l/\theta + \delta,
\end{equation}
where $\delta$ is the distance between the center of mass of the node and the geometric center (see fig.~\ref{opening_angles}). This criterion guarantees that if the center of mass is near the node's edge, only positions removed by an extra distance $\delta$ use the cell for a force evaluation. In case the center of mass is near the node's center, the old criterion (Eq.~\ref{THETA1}) is used. 

\subsection{The Parallel Tree-Algorithm}
\begin{figure}\centering
\begin{tikzpicture}
\node (start) [startstop] {Domain decomposition};
\node (move) [startstop, below of=start, yshift=-0.5cm,xshift=2cm] {Apply potential};
\node (construct) [startstop, below of=start, yshift=-0.5cm,xshift=-2cm] {Construct local trees};
\node (exchange) [startstop, below of=construct, yshift=-0.5cm,] {Exchange tree nodes};
\node (let) [startstop, below of=start, yshift=-3.5cm] {Build LET};
\node (walk) [startstop, below of=move, yshift=-0.5cm] {Tree walk};
\draw [arrow]  (start) to (construct);
\draw [arrow]  (construct) to (exchange);
\draw [arrow]  (exchange) to (let);
\draw [arrow]  (let) to (walk);
\draw [arrow]  (walk) to (move);
\draw [arrow]  (move) to (start);
\end{tikzpicture}
\caption{Flowchart of the parallel Barnes \& Hut algorithm as it is implemented with our GPU accelerated gravity solver into the FLASH code. Given an initial domain decomposition, we construct a BH tree covering the locally stored data cells. The local trees are traversed and essential tree nodes are exchanged between the processes. With the Essential Nodes and the local data, we build local essential tree (LET) and traverse it to calculate the gravitational potential. The calculated potential is then ready to use by other FLASH modules and the gravity solver can be called in the subsequent simulation step.}
\label{parallel_hut}
\end{figure}
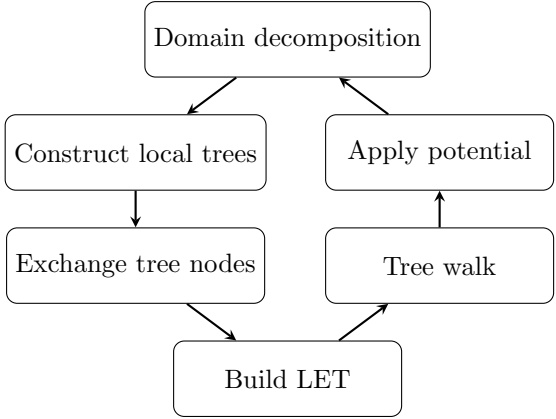

The parallelization of the  Barnes \& Hut  algorithm is not obvious, since the inhomogeneous distribution of the particles in the tree does not lead to simple load balanced domain decompositions. To distribute the work to many independent processors, the particles and the tree must be divided in a balanced way. Possible parallelization methods are described in \citep{Salmon:1991wb} and \citep{Dubinski:1996gr}. The basic workflow of one parallel Barnes \& Hut algorithm is outlined in fig.~\ref{parallel_hut}. The parallel  Barnes \& Hut algorithm starts with a domain decomposition to achieve a load balanced particle distribution. The decomposition can be carried out using any suitable algorithm e. g. the method of orthogonal recursive bisection the method of orthogonal recursive bisection \citep{Dubinski:1996gr}, the Morton space filling curve \citep{2012ASPC..453..325B} or a Peano-Hilbert space filling curve \citep{2014arXiv1412.0659B}.\footnote{The FLASH code implements the Morton space filling curve for domain decomposition.}
Given a load-balanced distribution of particles among different processors, a BH-tree can be constructed using the locally stored set of particles (fig.~\ref{parallel_hut}, Construct local trees). If every processor had a copy of the complete domain, the force calculation could be carried out now. Unfortunately, the amount of memory required for a full copy of the domain data is prohibitive and thus,  we use a smaller but sufficiently large enough sub-tree.\\
The \emph{LET (locally essential trees)} approach assumes that only a subset of tree nodes is necessary since the BH-tree can be pruned by using the opening angle criterion \citep{Salmon:1991wb}. Applying the opening angle criterion to the entire group of particles on a donating processor, a subset of nodes can be selected that are necessary for a successful force evaluation on a different processor. The selected nodes are then sent to the target processor (fig.~\ref{parallel_hut}, Exchange tree nodes). On the target processor, the existing local tree is enlarged in this way by a pruned version of the trees resident on other processors (fig.~\ref{parallel_hut}, Exchange tree nodes). This \emph{local essential tree} is then traversed and the gravitational potential for the particles is evaluated (fig.~\ref{parallel_hut}, Tree walk).

\section{The GPU Barnes-Hut Tree Code for FLASH4}
The approaches from \citep{Salmon:1991wb} and from \citep{Dubinski:1996gr} have been modified and implemented into the FLASH software \citep{2000ApJS..131..273F}. Our implementation of the  BH-tree algorithm is written in FORTRAN and C  for the code running on CPU's and in CUDA-C for the code running on the GPU-Device. Accessing the global data fields of the FLASH package and calling message-passing routines is solely done using CPU code. The tree walks and tree builds are carried out on the GPU-Devices using several CUDA kernels. 
\subsection{The CPU Code}
\begin{figure}[]
\begin{tikzpicture}
\node (flash) [flash] {Flash 4};
\node (collect) [process, right of=start, xshift=4.5cm,yshift=8.5cm ] {Collect cell data (CPU)};
\node (exchange) [process, below of=collect,yshift=-0.8cm] {Exchange basic data (CPU, MPI)};
\node (cal_and_send) [process, below of=exchange, yshift=-0.8cm,text width=4cm] {Calculate and exchange Essential Nodes (CPU, GPU, MPI)};
\node (construct) [process, below of=cal_and_send, yshift=-1.0cm,text width=4cm] {Construct BH-LET and calculate potential (GPU)};
\node (fin) [process, below of=construct, yshift=-0.8cm] {Write forces to solution vector (CPU)};

\draw [arrow]  (1,8.5) to (collect);
\draw [arrow]  (collect) to (exchange);
\draw [arrow]  (exchange) to (cal_and_send);
\draw [arrow]  (cal_and_send) to  (construct);
\draw [arrow]  (construct) to (fin);
\draw [arrow]  (fin) to (1,1);
\end{tikzpicture}
\caption{Basic outline of the algorithm steps used in in the GPU-BH tree code. Notes in brackets mark whether  the respective routine runs on the CPU or the GPU and if message-passing routines are called (MPI). }
\label{GPU_BH_STEPS}
\end{figure}
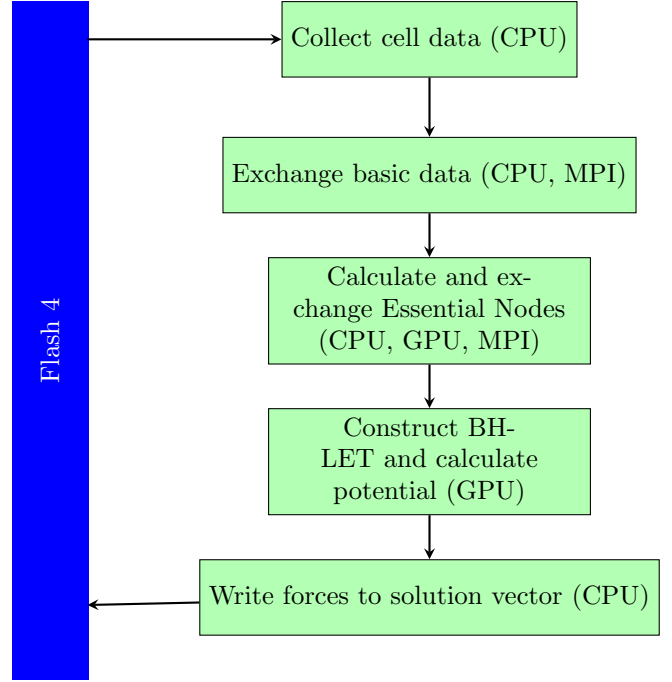

The major part of the CPU Code covers the allocation of data fields as well as accessing the FLASH routines for writing and reading global fields. Performance wise, the most important parts of the CPU code are the calls to message-passing routines. Generally, message-passing takes a large amount of time and therefore plays an important role in optimization procedures. In our implementation, message-passing routines are called to communicate pruned BH-trees and BH root node data. 
The naive strategy to calculate the forces and communicate data-on-demand requires a two-way communication and may cause a large communication overhead if not programmed carefully. Furthermore, a data-on-demand communication does not follow the aim to calculate the forces on the GPU. A GPU kernel would have to be stopped while waiting for the data and restarted after the data was received. To overcome this limitation, the  BH-LET can be constructed prior to the force calculation \citep{Liu:2000hm}. For each FLASH call to calculate the gravitational potential, every process runs through a series of steps. Figure~\ref{GPU_BH_STEPS} illustrates a basic flowchart of the algorithm.\newline 
Whenever the GPU-BH tree solver is called from inside the FLASH4 environment, the AMR-Nodes are expanded and the cell data is read (fig.~\ref{GPU_BH_STEPS}, Collect cell data). As a next step, some basic information about the domain's size and its orientation is distributed along all processes (fig.~\ref{GPU_BH_STEPS}, Exchange basic data). In step three (fig.~\ref{GPU_BH_STEPS}, Calculate and exchange Essential Nodes), we build the local BH-trees and calculate the LET nodes for each process, which are delivered to the respective processes. The tree builds and the tree walks in this step are implemented solely on the GPU device. In step four (fig.~\ref{GPU_BH_STEPS}, Construct BH-LET and calculate potential), we build the final BH-LET on the GPU-device and calculate the gravitational potential. Finally, the calculated potential is written back to the FLASH internal solution vector (fig.~\ref{GPU_BH_STEPS}, Write forces to solution vector).

\subsubsection{Collect cell data and exchange basic data}
The FLASH4 simulation software \citep{2000ApJS..131..273F} uses the PARAMESH library \citep{2000CoPhC.126..330M,Plewa:2005wj,Deane:2006tj} to handle its AMR mesh and the domain decomposition. Whenever the GPU-BH tree solver is called from inside the FLASH4 environment, the AMR-Nodes are expanded and the AMR cell data is read (fig.~\ref{GPU_BH_STEPS}, Collect cell data). To construct a local root node, the Center of Mass of the local data cells is calculated on the fly while parsing the AMR tree. Position data and the respective mass of every data cell is read and stored in arrays with a one-dimensional layout. For later usage on the GPU, the arrays are structured into blocks holding the data of exactly 32 neighboring cells. This does not only preserve locality of the data cells but matches the $warp$ \footnote{A $warp$ is a set of threads scheduled at the same time on the CUDA device. The maximum number of threads in a warp is 32.} size of the GPU, which is beneficial for the tree walks.\\
A virtual root node (in the following referred to as ``VN'') located at the center of mass of all local data cells is constructed. Every VN holds position and mass data as well as its size in a double precision floating point variable. Because every process needs a copy of all VNs and the amount of data communicated in this step is comparably small, a single {\tt{MPI\_Allgather}} call realizes the VN exchange.\footnote{Using {\tt{MPI\_Allgather}}, the block of data sent from the jth process is received by every process and placed in the jth block of the receive buffer.} This way, the owner of every VN can be figured out solely by the index. 

\subsubsection{Calculate and exchange Essential Nodes}\label{Calc_send}
\begin{figure}[]
\begin{tikzpicture}[auto, node distance=1.5cm,->,>=stealth,shorten >=1pt]
\tikzstyle{vspecies}=[rectangle, minimum size=0.5cm,text width=1cm,draw=blue!80,fill=blue!20]
\tikzstyle{square}=[rectangle,thick,minimum size=0.5cm,draw=red!80,text width=1cm,fill=green!20]
\tikzstyle{fspecies}=[circle, minimum size=0.5cm,draw=red!80,fill=red!20]

\node [fspecies] (P1) {P1};
\node [fspecies, below of = P1, yshift=-1cm] (PN) {P2};
\node [fspecies, below of = PN, yshift=-1cm] (PX) {PN};

\node [vspecies, right of = PN,xshift=5.5cm] (BT2) {Build tree};

\node [vspecies, right of = P1] (BT) {Build tree};
\node [vspecies, right of = BT] (VN) {Walk tree};
\node [above right of = VN,xshift = 0.25cm] (CN) {For each VN};
\node [square, right of = VN,xshift = 0.8cm] (SN) {Send Nodes};

\draw [line width = 1pt,dotted,red] (P1) -- (PN);
\draw [line width = 1pt,dotted,red] (PN) -- (PX);

\draw [line width = 1pt] (P1) -- (BT);
\draw [line width = 1pt] (BT) -- (VN);
\draw [line width = 1pt] (PN) -- (BT2);
\draw [line width = 1pt] (BT2) -- (8,-2.5);
\draw [line width = 1pt] (PX) -- (8,-5);
\draw [line width = 1pt] (5.6,-0.448) -- (5.6,-5);
\draw [line width = 1pt] (5.2,-0.448) -- (5.2,-2.5);
\draw [line width = 1pt] (SN) -- (8,0);

 \path[every node/.style={font=\sffamily\small}]
    (VN)  edge [bend left] node[left] {} (SN)
    (SN)  edge [bend left] node[right] {} (VN);

\end{tikzpicture}
\caption{Flowchart of the steps carried out when calculating and distributing Essential Nodes. Blue boxes indicate procedures carried out on the GPU device, green boxes indicate CPU procedures. A serial GPU-device access is assumed for all contributing processes $P1...PN$ (red circles). Process $P1$ acquires the GPU and starts building a local BH tree. Other processes ($P2...PN$) have to wait until the GPU device is freed by $P1$. The local BH tree is traversed for each imported VN (blue box (Walk tree)) and the calculated Essential Nodes are sent to the respective processes (green box (send Nodes)). In case all VNs are processed, $P1$ frees the GPU device and P2 can acquire it to fulfill the same tasks.}
\label{Calc_Ex_Fig}
\end{figure}
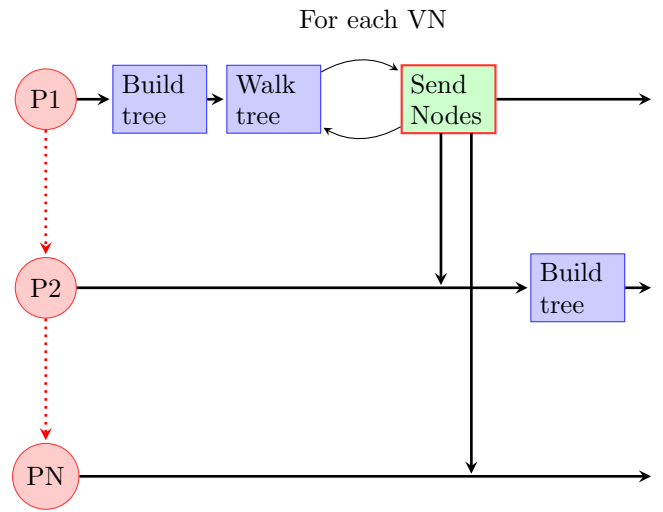

As mentioned before, the overlap of communication time with calculation is an important concept in distributed memory approaches. When using the LET approach, the chance of a communication and computation overlap during the force evaluation is obviously lost. Since the vast majority of communication time is needed when the Essential Nodes are exchanged, the respective calculations can take place during this communication procedure (fig.~\ref{GPU_BH_STEPS}, Calculate and send essential sub-trees). \\
Figure~\ref{Calc_Ex_Fig} shows a flowchart of the procedures and how the overlap is implemented. Whenever a process acquires a GPU-device, the device is blocked for other processes and a local BH tree is build. The tree is traversed on the GPU for every imported VN to calculate the respective Essential Nodes. After all Essential Nodes for one VN are calculated; an index array is  copied to host memory\footnote{Host memory is maintained by the CPU in its own separate memory space in DRAM}.
We use an index array instead of the floating-point node data to reduce the memory operations during the main loop. The actual array of the node data is based on the index array and constructed on the CPU and is sent to the respective VN owner using non-blocking routines.
The message size varies depending on the cell opening criterion and the number of data cells a processor holds. For a process holding n data cells, the number of Essential Nodes may vary between 1 (the process just sends its root node) and  n (the process sends all data cells). Processes holding no data cells do not access the GPU device and do not build up a local tree. Figure~\ref{ESS_ARRAYS} illustrates the structure of the index- and the exchange-array.\newline
\begin{figure}[]
\centering
\begin{tikzpicture}
\node at (2.8,6.8) {Essentials index:};

\filldraw[draw=black,fill=green!50] (1.5,6) rectangle (6,6.5);
\draw (2,6) -- (2,6.5);
\draw (2.5,6) -- (2.5,6.5);
\draw (3,6) -- (3,6.5);
\draw (3.5,6) -- (3.5,6.5);

\node at (1.75,6.25) {$P_1$};
\node at (2.25,6.25) {$P_2$};
\node at (2.75,6.25) {$P_3$};
\node at (3.25,6.25) {$P_4$};
\node at (3.75,6.25) {$P_5$};
\draw (4,6) -- (4,6.5);
\node at (4.75,6.25) {...};
\draw (5.5,6) -- (5.5,6.5);
\node at (5.75,6.25) {$P_N$};

\node at (2.8,4.8) {Exchange array:};
\filldraw[draw=black,fill=lightgray] (1.5,4) rectangle (9.5,4.5);
\node at (1.75,4.25) {$x_1$};
\draw (2,4) -- (2,4.5);
\node at (2.5,4.25) {...};
\draw (3,4) -- (3,4.5);
\node at (3.25,4.25) {$x_N$};
\draw (3.5,4) -- (3.5,4.5);
\node at (3.75,4.25) {$y_1$};
\draw (4,4) -- (4,4.5);
\node at (4.5,4.25) {...};
\draw (5,4) -- (5,4.5);
\node at (5.25,4.25) {$y_N$};
\draw (5.5,4) -- (5.5,4.5);
\node at (5.75,4.25) {$z_1$};
\draw (6,4) -- (6,4.5);
\node at (6.5,4.25) {...};
\draw (7,4) -- (7,4.5);
\node at (7.25,4.25) {$z_N$};
\draw (7.5,4) -- (7.5,4.5);
\node at (7.75,4.25) {$m_1$};
\draw (8,4) -- (8,4.5);
\node at (8.5,4.25) {...};
\draw (9,4) -- (9,4.5);
\node at (9.25,4.25) {$m_N$};
\end{tikzpicture}
\caption{Structure of the index array and the exchange array. The index array holds pointers to the position and mass data of Essential Nodes and data cells. For $N$ Essential Nodes, $N$ integer values are copied from device.memory to host-memory. The exchange array is built by means of the index array holding $4N$ entries. }
\label{ESS_ARRAYS}
\end{figure}
The next VN can be processed immediately, because non-blocking MPI routines ({\tt{MPI\_ISend}}) are used for the communication in this step. On the receiver's side, the Essential Nodes are received while waiting for a free GPU device. Ideally, the waiting times to access a GPU device and the actual calculation time on the device overlap with the communication time. This is true for all but the first processes accessing the GPU device. The actual procedures carried out on the GPU device are outlined in section~\ref{kernels}.

\subsection{The CUDA Kernels}\label{kernels}
The whole tree building process together with the calculation of the essential sub trees and the gravitational potentials is solely done in device memory. This approach minimizes memory transactions like copying data from the host to the device or vice verse. Only position and mass data is copied to the device and just results are copied back to host memory. The construction of the  Barnes \& Hut trees and the memory layout is based on the findings reported by \citep{Burtscher:2011ui}. In the following, the data layout and the largest CUDA kernels are described in detail. 

\subsubsection{Data Layout}
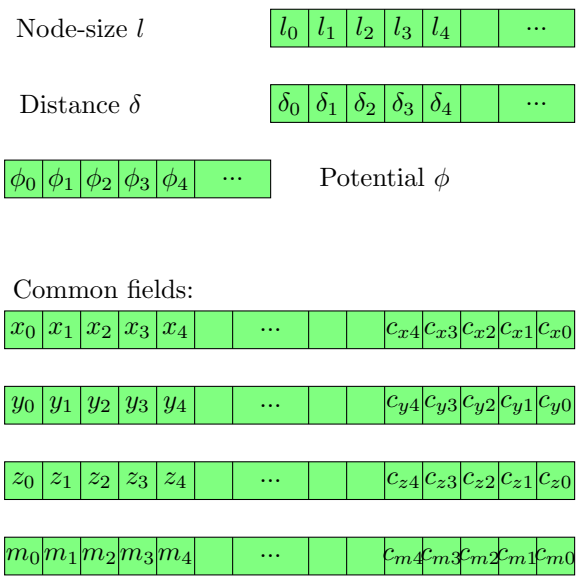
\begin{figure}[]
\centering
\begin{tikzpicture}

\node at (2.5,10.25) {Node-size $l$};
\filldraw[draw=black,fill=green!50] (5,10) rectangle (9,10.5);
\node at (8.5,10.25) {...};
\draw (5.5,10) -- (5.5,10.5);
\draw (6,10) -- (6,10.5);
\draw (6.5,10) -- (6.5,10.5);
\draw (7,10) -- (7,10.5);
\draw (7.5,10) -- (7.5,10.5);
\draw (8,10) -- (8,10.5);

\node at (5.25,10.25) {$l_0$};
\node at (5.75,10.25) {$l_1$};
\node at (6.25,10.25) {$l_2$};
\node at (6.75,10.25) {$l_3$};
\node at (7.25,10.25) {$l_4$};

\node at (2.5,9.25) {Distance  $\delta$};
\filldraw[draw=black,fill=green!50] (5,9) rectangle (9,9.5);
\node at (8.5,9.25) {...};

\draw (5.5,9) -- (5.5,9.5);
\draw (6,9) -- (6,9.5);
\draw (6.5,9) -- (6.5,9.5);
\draw (7,9) -- (7,9.5);
\draw (7.5,9) -- (7.5,9.5);
\draw (8,9) -- (8,9.5);

\node at (5.25,9.25) {$\delta_0$};
\node at (5.75,9.25) {$\delta_1$};
\node at (6.25,9.25) {$\delta_2$};
\node at (6.75,9.25) {$\delta_3$};
\node at (7.25,9.25) {$\delta_4$};

\node at (6.5,8.25) {Potential  $\phi$};
\filldraw[draw=black,fill=green!50] (1.5,8) rectangle (5,8.5);
\draw (2,8) -- (2,8.5);
\draw (2.5,8) -- (2.5,8.5);
\draw (3,8) -- (3,8.5);
\draw (3.5,8) -- (3.5,8.5);

\node at (1.75,8.25) {$\phi_0$};
\node at (2.25,8.25) {$\phi_1$};
\node at (2.75,8.25) {$\phi_2$};
\node at (3.25,8.25) {$\phi_3$};
\node at (3.75,8.25) {$\phi_4$};
\draw (4,8) -- (4,8.5);
\node at (4.5,8.25) {...};

\node at (2.8,6.8) {Common fields:};
\filldraw[draw=black,fill=green!50] (1.5,6) rectangle (9,6.5);
\draw (2,6) -- (2,6.5);
\draw (2.5,6) -- (2.5,6.5);
\draw (3,6) -- (3,6.5);
\draw (3.5,6) -- (3.5,6.5);

\node at (1.75,6.25) {$x_0$};
\node at (2.25,6.25) {$x_1$};
\node at (2.75,6.25) {$x_2$};
\node at (3.25,6.25) {$x_3$};
\node at (3.75,6.25) {$x_4$};
\draw (4,6) -- (4,6.5);
\draw (4.5,6) -- (4.5,6.5);
\node at (5,6.25) {...};
\draw (5.5,6) -- (5.5,6.5);
\draw (6,6) -- (6,6.5);
\draw (6.5,6) -- (6.5,6.5);
\node at (6.75,6.25) {$c_{x4}$};
\draw (7,6) -- (7,6.5);
\node at (7.25,6.25) {$c_{x3}$};
\draw (7.5,6) -- (7.5,6.5);
\node at (7.75,6.25) {$c_{x2}$};
\draw (8,6) -- (8,6.5);
\node at (8.25,6.25) {$c_{x1}$};
\draw (8.5,6) -- (8.5,6.5);
\node at (8.75,6.25) {$c_{x0}$};

\filldraw[draw=black,fill=green!50] (1.5,5) rectangle (9,5.5);
\draw (2,5) -- (2,5.5);
\draw (2.5,5) -- (2.5,5.5);
\draw (3,5) -- (3,5.5);
\draw (3.5,5) -- (3.5,5.5);

\node at (1.75,5.25) {$y_0$};
\node at (2.25,5.25) {$y_1$};
\node at (2.75,5.25) {$y_2$};
\node at (3.25,5.25) {$y_3$};
\node at (3.75,5.25) {$y_4$};
\draw (4,5) -- (4,5.5);
\draw (4.5,5) -- (4.5,5.5);
\node at (5,5.25) {...};
\draw (5.5,5) -- (5.5,5.5);
\draw (6,5) -- (6,5.5);
\draw (6.5,5) -- (6.5,5.5);
\node at (6.75,5.25) {$c_{y4}$};
\draw (7,5) -- (7,5.5);
\node at (7.25,5.25) {$c_{y3}$};
\draw (7.5,5) -- (7.5,5.5);
\node at (7.75,5.25) {$c_{y2}$};
\draw (8,5) -- (8,5.5);
\node at (8.25,5.25) {$c_{y1}$};
\draw (8.5,5) -- (8.5,5.5);
\node at (8.75,5.25) {$c_{y0}$};

\filldraw[draw=black,fill=green!50] (1.5,4) rectangle (9,4.5);
\draw (2,4) -- (2,4.5);
\draw (2.5,4) -- (2.5,4.5);
\draw (3,4) -- (3,4.5);
\draw (3.5,4) -- (3.5,4.5);

\node at (1.75,4.25) {$z_0$};
\node at (2.25,4.25) {$z_1$};
\node at (2.75,4.25) {$z_2$};
\node at (3.25,4.25) {$z_3$};
\node at (3.75,4.25) {$z_4$};
\draw (4,4) -- (4,4.5);
\draw (4.5,4) -- (4.5,4.5);
\node at (5,4.25) {...};
\draw (5.5,4) -- (5.5,4.5);
\draw (6,4) -- (6,4.5);
\draw (6.5,4) -- (6.5,4.5);
\node at (6.75,4.25) {$c_{z4}$};
\draw (7,4) -- (7,4.5);
\node at (7.25,4.25) {$c_{z3}$};
\draw (7.5,4) -- (7.5,4.5);
\node at (7.75,4.25) {$c_{z2}$};
\draw (8,4) -- (8,4.5);
\node at (8.25,4.25) {$c_{z1}$};
\draw (8.5,4) -- (8.5,4.5);
\node at (8.75,4.25) {$c_{z0}$};

\filldraw[draw=black,fill=green!50] (1.5,3) rectangle (9,3.5);
\draw (2,3) -- (2,3.5);
\draw (2.5,3) -- (2.5,3.5);
\draw (3,3) -- (3,3.5);
\draw (3.5,3) -- (3.5,3.5);

\node at (1.75,3.25) {$m_0$};
\node at (2.25,3.25) {$m_1$};
\node at (2.75,3.25) {$m_2$};
\node at (3.25,3.25) {$m_3$};
\node at (3.75,3.25) {$m_4$};
\draw (4,3) -- (4,3.5);
\draw (4.5,3) -- (4.5,3.5);
\node at (5,3.25) {...};
\draw (5.5,3) -- (5.5,3.5);
\draw (6,3) -- (6,3.5);
\draw (6.5,3) -- (6.5,3.5);
\node at (6.75,3.25) {$c_{m4}$};
\draw (7,3) -- (7,3.5);
\node at (7.25,3.25) {$c_{m3}$};
\draw (7.5,3) -- (7.5,3.5);
\node at (7.75,3.25) {$c_{m2}$};
\draw (8,3) -- (8,3.5);
\node at (8.25,3.25) {$c_{m1}$};
\draw (8.5,3) -- (8.5,3.5);
\node at (8.75,3.25) {$c_{m0}$};


\end{tikzpicture}
\caption{Illustration of the device array structures. The Node-size $l$ and the the distance $delta$ is only used for internal nodes. The array for the gravitational potential is only used for leaf nodes (data cells). Position ($x, y, z$) and mass ($m$) arrays are used for both.}
\label{ARRAY_STRUCTS}
\end{figure}

We use several aligned scalar arrays and array indices to allow for coalesced memory access. The common fields (three dimensional position and mass) for cell data (leaf nodes) and internal nodes are represented with four floating-point arrays. The leaves are allocated at the beginning and the internal nodes at the end of the arrays. Other fields are only valid for leaf nodes or internal nodes. For the latter, we use one array to store the node size ($l$) and one array to store the distance ($\delta$) between the node's center of mass and its geometric center. The node size $l$ is calculated and stored during the tree build (see section:~\ref{TreeBuild}) and the distance $\delta$ is evaluated while calculating the centers of mass (see section:~\ref{CenterGrav}). Finally, the array for the gravitational potential is only used for the leaf nodes. 
Figure~\ref{ARRAY_STRUCTS} outlines the described memory structure.

\subsubsection{Tree Building Kernel}\label{TreeBuild}
The \emph{Tree Building Kernel} implements an iterative tree-building algorithm. Starting with a root node, all local data cells and all Essential Nodes are inserted into the tree. In case of a local BH-tree, as it is used for determining Essential Nodes, the root node's position data exactly at the geometric center of all local data cells. In case of a LET, the root node is located at the domain's center. For the LET it is important that all processes agree on the root node level, because Essential Nodes are treated as data cells during the tree build. 
In the \emph{Tree Building Kernel}, data cells are assigned to blocks and threads within blocks in a round robin fashion\footnote{CUDA blocks are organized into a one-dimensional, two-dimensional, or three-dimensional grid of thread blocks. On current GPUs, a thread block may contain up to 1024 threads \citep{NVIDIA:tc}.}. Every thread traverses the tree down to the desired leaf node and inserts an index pointing to its cell into the free leaf node. For this, the respective leaf node is locked such that other threads attempting to access this node have to wait until the appropriate index is inserted. 

\subsubsection{Centers of Gravitation Kernel}
\label{CenterGrav}
In the \emph{Centers of Gravitation Kernel} the masses of the internal nodes are calculated, their three-dimensional position is updated to the center of gravity and the distance between the node's center of gravity and its geometric center is evaluated. During the tree build all internal nodes are initiated with a negative mass indicating that their actual mass needs to be calculated. Since mass data is only available for the leaf nodes (the nodes containing cell data), the masses are summed up in a bottom-up manner. The tree is walked from bottom up where internal nodes are assigned to blocks and threads within blocks. Here, every thread processes one node and its direct children. For a node $N$ with $n$ children $C_\mathrm{i}, \mathrm{i}=1...\mathrm{n}$ each with mass $m_\mathrm{i}$ and position data with coordinates $r_\mathrm{i}, \mathrm{i}=1,..,\mathrm{n}$. The mass of the node $M$ is simply the sum of the children's masses $m_\mathrm{i},..,m_\mathrm{n}$ and its position is at $R$ with:
\begin{equation}R=\frac{1}{M}\sum^n_{\mathrm{i}=1}m_\mathrm{i} r_\mathrm{i}\end{equation}
A thread with its assigned internal node at the very bottom of the tree simply sums up the masses and updates its node mass. The geometric center and its distance to $R$ is calculated and stored in a separate array before the node's mass and position is updated. Threads that encounter a child with negative mass wait until its mass is updated by another thread.

\subsubsection{Find Essential Nodes Kernel}
\begin{figure}[]
\centering
\begin{tikzpicture}
\draw (0,0) -- (0, 2.3);
\draw (0,0) -- (2.3, 0);
\draw (0,2.3) -- (2.3, 2.3);
\draw (2.3,0) -- (2.3, 2.3);

\draw (4,0) -- (4, 4);
\draw (4,4) -- (8, 4);
\draw (4,0) -- (8, 0);
\draw (8,0) -- (8, 4);

\node (CENT) at (6,2) [circle,draw,shade,minimum size=0.15] {};
\node (COM) at (5.6,0.7) [circle,draw,shade,minimum size=0.15] {};

\node[right of = COM , yshift = -0.3cm,xshift=-05] {C.O.M};
\node at (2,3) [] {\large $d_c > l/\theta +\delta$};

\draw[<->, very thick] (4,3.5) to node[above] {$l$} (8,3.5);

\draw[<->, very thick] (2.3,0.7) to node[below, xshift=-0.2] {$d_c$} (COM);
\draw[<->, very thick] (CENT) to node[right] {$\delta$} (COM);

\end{tikzpicture}
\caption{The geometry of the modified opening criterion used in the GPU-BH tree code when calculating the Essential Nodes.}
\label{opening_angles2}
\end{figure}
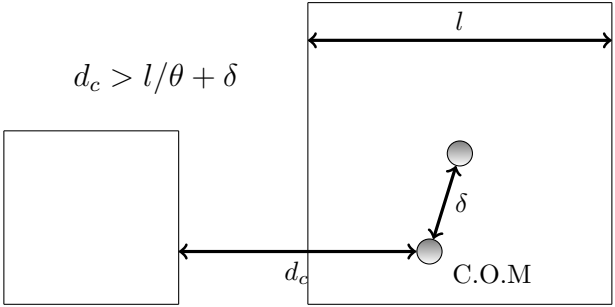

To generate the essential sub trees, the local  BH-trees must be traversed with regard to the VNs. Nodes and data cells are treated as essential if they pass the opening angel test. We use a slightly modified version of the opening angle criterion described in Eq.~\ref{THETA2}. The imported VNs are treated as nodes during the tree walk, and the distance between a VN and an internal BH node must be evaluated. Here, we refer to the minimum distance $d_\mathrm{c}$ between the nodes center of mass and the VN (see fig.~\ref{opening_angles2}). \newline
In general, the selection of Essential Nodes can be done using any tree traversal method. A simple algorithm targeted to GPU devices would be to transform a depth-first or breath-first traversal with each VN assigned to one GPU-Thread. To support the earlier described method of overlapping and to minimize the memory footprint, the selection of Essential Nodes is implemented for single VNs. Opposed to assigning the VNs to threads, the local tree nodes are assigned to blocks and threads within blocks in a round robin fashion starting from the root node. The \emph{Find Essential BH-Nodes Kernel} generates an index list holding indices for all nodes and data cells treated as essential for the processed VN. The CUDA kernel is called once for each depth and thus block level synchronization is assured. Neither atomic operations nor direct synchronization calls are used. The generated index list is copied back to host memory, where the respective communication buffer is filled with the position and mass data of the Essential Nodes and data cells.

\subsection{Calculate Gravitational Potential Kernel}
Given a tree with updated centers of gravity and masses, the gravitational potential is calculated for all local data cells. The corresponding CUDA-Kernel performs a stack based depth-first tree traversal. The stack is created in shared memory to guarantee fast memory access. The stack size is aligned to the warp size and only the first thread of a warp (lane 0) is allowed to modify the stack. Local data cells are assigned to blocks and threads within blocks in a round robin fashion. The first thread of the warp traverses the tree and pushes the nodes onto the stack. Every thread reads the top node on the stack and calculates the distance between its data cell and the node in the stack. The opening angle parameter test according to Eq.~\ref{THETA2} is applied and the gravitational potential for the data cell is updated in cache. 
Note that the opening angle test works as a data-dependent conditional branch where threads in a warp could diverge. This is a potential performance bottleneck, because the warp would serially execute each taken branch. To overcome this performance lack, the thread voting function {\tt{all()}}\footnote{The {\tt{all()}} warp voting function evaluates a predicate for all active threads of the warp and returns non-zero if and only if the predicate evaluates to non-zero for all of them \citep{NVIDIA:tc}.} is used for the outcome of the test. As a result, every thread in the warp ``opens'' a node even if just one of the threads in the warp has a respective test outcome. Although this prevents thread divergence and can potentially increases the accuracy, the beneficial effect of the opening angle test is obvious lost for some threads. In a worst-case scenario every warp traverses the whole tree because the warps data cells do not share the same test outcomes. The number of diverging test outcomes can be reduced if warps process only neighboring data cells, because nearby data cells share the same interaction list \citep{Barnes:1990tc}. For this, we store the data cells in blocks of exactly 32 (warp size) neighbors such that every warp reads a set of nearby data cells. 
 
\section{Results}\label{RESandDISC}
In this section, we describe the results of three astrophysical motivated test problems solved using our GPU-BH tree algorithm: the gravitational potential of a homogeneous spheroid, a homogeneous spheroid with initial velocity turbulence and a homogeneous oblate spheroid. All simulations were carried out in FLASH4.2.2 \citep{2000ApJS..131..273F}.\newline
The simulations  were carried out on our institute's cluster where every compute node holds two Intel Xeon Processors with each having six cores with a clock speed of 2.4 GHz and one NVIDIA Tesla C2075 GPU device with 6 GB of GDDR5 memory and 448 CUDA cores with a clock speed of 1.15 GHz. The CPU-Code was compiled using the Intel compilers (ICC v. 10.1, IFORT v. 10.1) with aggressive optimization flags (-O3). The GPU code was compiled using the NVCC compiler (Cuda compilation tools, release 7.0, V7.0.27) with default settings. The runtime of the simulations were measured using different timer functions provided by the FLASH4 API. 

\begin{figure*}[]
\subfigure{\includegraphics[width=0.5\textwidth]{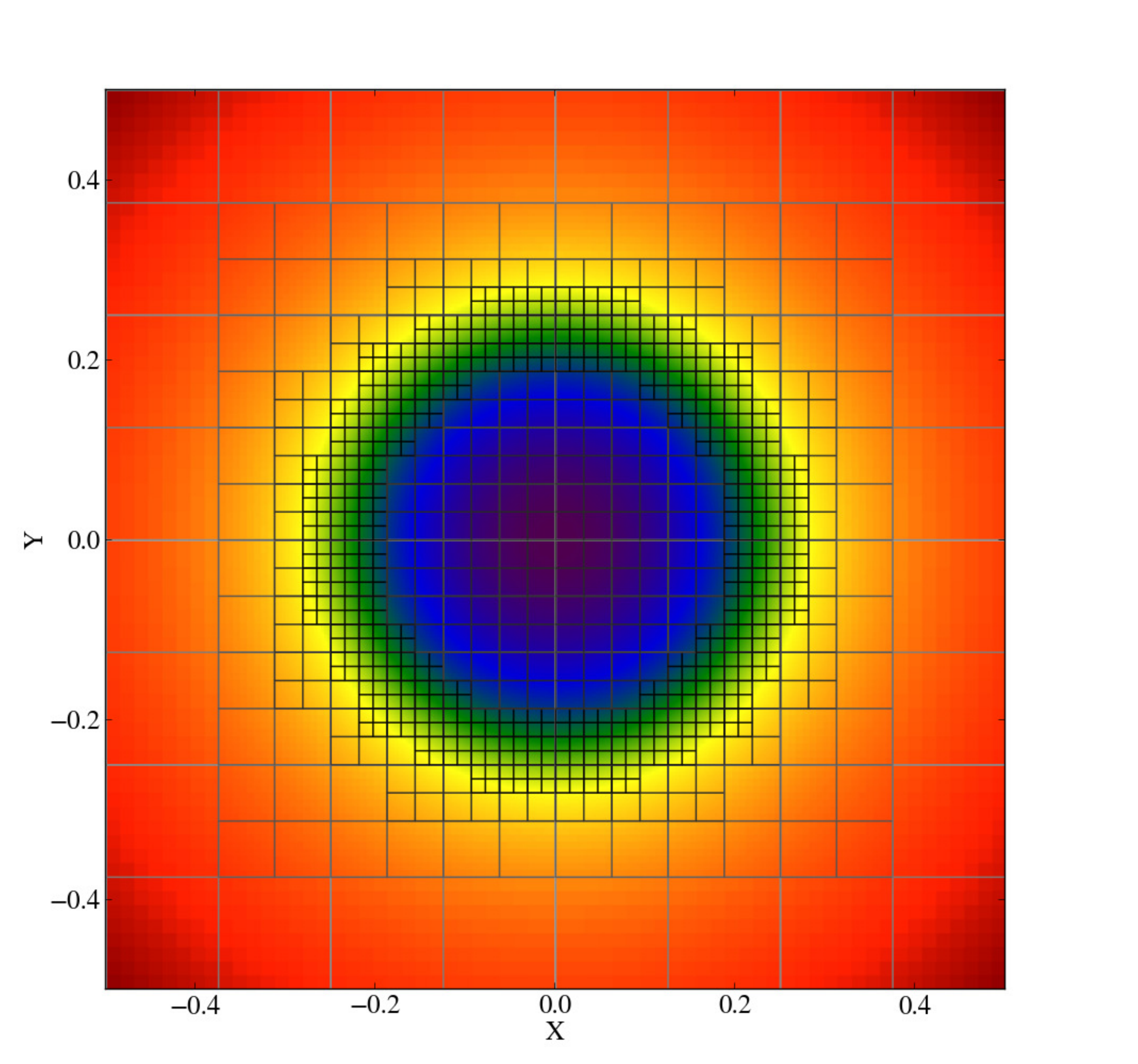}}\hfill
\subfigure{\includegraphics[width=0.5\textwidth]{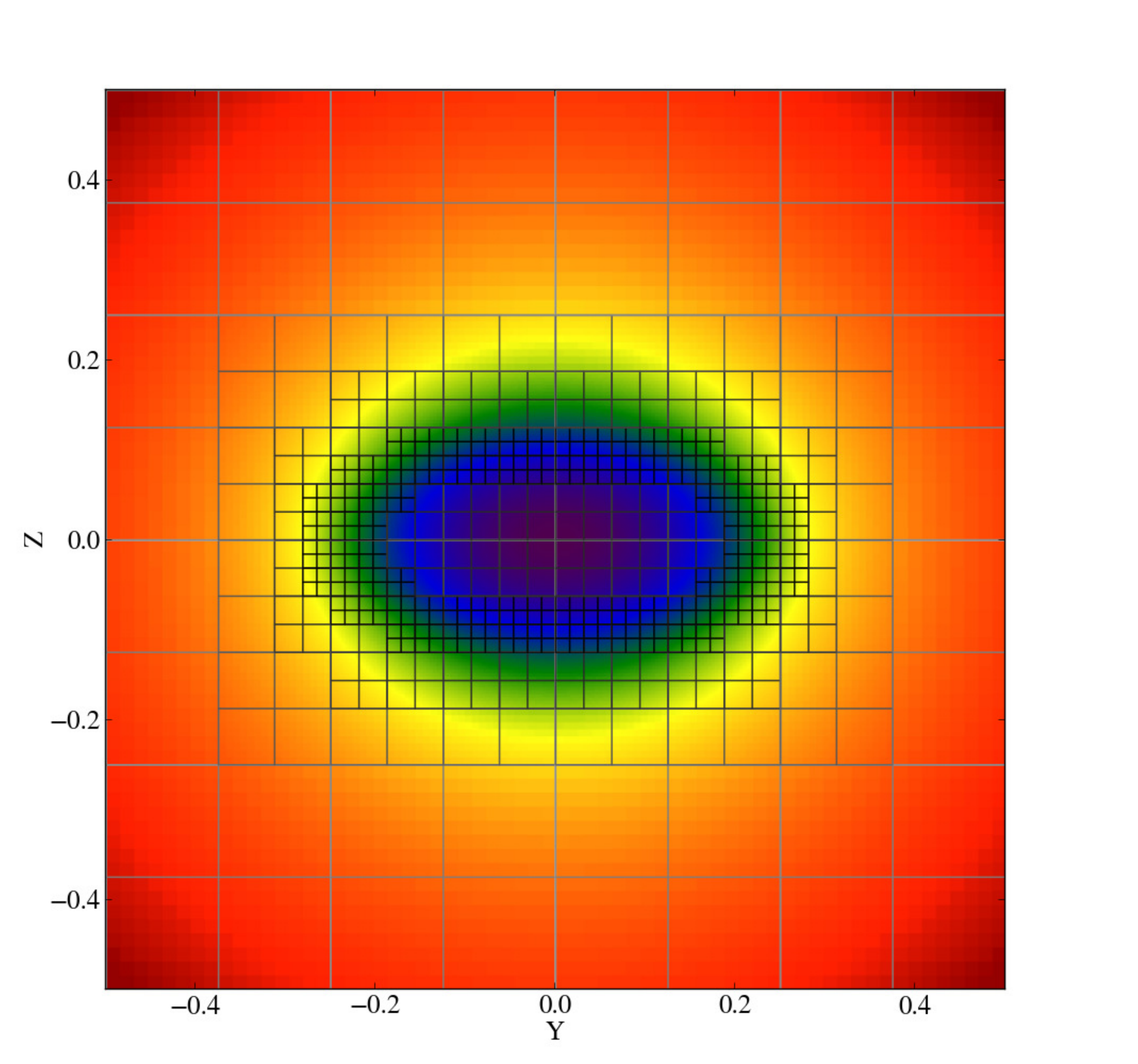}}
\caption{Potential of a MacLaurin spheroid with eccentricity 0.9 as computed with the GPU-BH tree code using $\alpha=0.25$ and using an adaptively refined mesh with seven levels of refinement. The left image shows the potential and AMR block outlines (grey and black lines) in the x-y plane passing through the center of the ellipsoid. The right image shows the same quantities for the y-z plane.}
\label{LAURIN_FIG}
\end{figure*}
\FloatBarrier
\subsection{Accuracy with the MacLaurin Sphere}
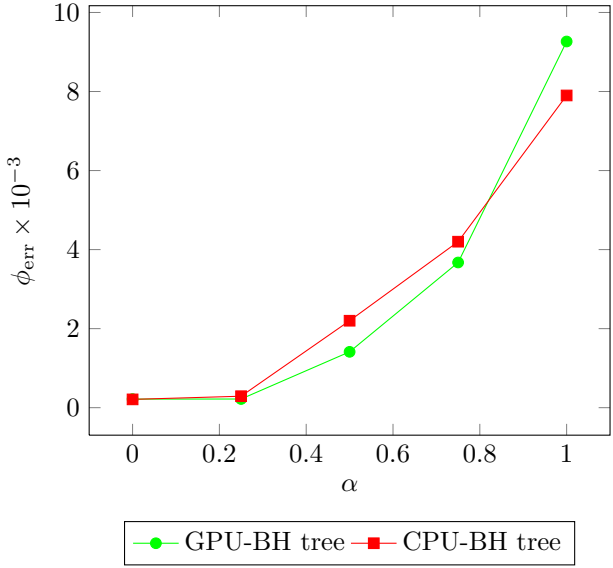
\begin{figure}[]
  \begin{tikzpicture}
    \begin{axis}[
    legend style={at={(0.5,-0.2)},
    anchor=north,legend columns=2},
     legend entries={GPU-BH tree,  CPU-BH tree},
      xlabel={$\alpha$},
      ylabel={$\phi_\mathrm{err} \times 10^{-3}$},
      scaled y ticks=false,
      xticklabel style={/pgf/number format/.cd,fixed},
      yticklabel style={/pgf/number format/.cd,fixed},
    ]
    
     \addplot [
     color=green, 
     mark = *] 
    table [
   x = CPU,
  y = SPUP
   ] {
     SPUP CPU
    9.265448373035489      1.0  
    3.673746141417680      0.75  
    1.413135457138840      0.5 
    0.2192557562353471      0.25
    0.2171458809279802      0.0 
};
\label{acc_gpu}

   \addplot [
color=red, 
mark = square*] 
table [
x = CPU,
y = SPUP
   ] {
     SPUP CPU
     7.90      1.0  
     4.20      0.75  
     2.20      0.5 
     0.29    0.25
     0.21    0.0 
};
\label{acc_cpu}

    \end{axis}

  \end{tikzpicture}
\caption{Maximum relative errors ($\phi_\mathrm{err}$) in the gravitational potential for the MacLaurin sphere calculated with the GPU-BH tree code (\ref{acc_gpu}) and the CPU-BH tree code (\ref{acc_cpu}) for different  $\alpha$ values. The GPU-tree code produces a higher accuracy for $\alpha$ values between $0.2$ and $0.8$ while the CPU tree code shows an accuracy advantage for $\alpha$ values $> 0.8$. Here, an alpha value of $0.0$ refers to the direct $O(n^2)$ algorithm.}
\label{Errors}
\end{figure}

To test the accuracy and the influence of the opening angle parameter $\alpha$, we use a so called ``MacLaurin'' spheroid. The gravitational potential at the surface of, and inside such a homogeneous spheroid is expressible in terms of analytical functions \citep{Chandrasekhar:1969uh}.\\
For a point inside the spheroid with density $\rho$ the gravitational potential is
\begin{equation}
\Phi(x) =\pi G \rho [2A_1 a^2_1 - A_1(x^2+y^2) - A_3(a_3^2 -z^2)] ,
\end{equation}
where $a_1$, $a_2$ and $a_3$ are the semi major axes of the spheroid and $a_1= a_2 > a_3$. Here 
\begin{equation}
A_1 = \frac{\sqrt{1-e^2}}{e^3}\sin^{-1}e-\frac{1-e^2}{e^2},
\end{equation}
\begin{equation}
A_3 = \frac{2}{e^2}-\frac{2\sqrt{1-e^2}}{e^3}\sin^{-1}e,
\end{equation}
where e is the ellipticity of the spheroid:
\begin{equation}
e=\sqrt{1-\Big(\frac{a_3}{a_2}\Big)^2}.
\end{equation}
For a point outside the spheroid, the potential is:
\begin{equation}
\begin{split}
\Phi(x)& =\frac{2a_3}{e^2}\pi G\rho\\ 
&\times \Bigg[ a_1e \tan^{-1}h- \frac{1}{2}\Bigg((x^2+y^2)\Bigg(\tan^{-1}h-\frac{h}{1+h^2}\Bigg)\\ 
& + 2z^2(h-\tan^{-1}h)\Bigg)\Bigg],
\end{split}
\end{equation}
where
\begin{equation}
h = \frac{a_1e}{\sqrt{a_3^2 + \lambda}},
\end{equation}
and $\lambda$ is the positive root of the equation
\begin{equation}
\begin{split}
\frac{x^2}{a_1^2 +\lambda} + \frac{y^2}{a_2^2 +\lambda} + \frac{z^2}{a_3^2 +\lambda} = 1.
\end{split}
\end{equation}
The simulation was setup with a uniform density $p=1\ \mathrm{g\ cm}^{-3}$ inside the spheroid and $p = \epsilon \rightarrow 0$ outside the spheroid with an eccentricity of $0.9$. The spheroids were centered in a box with unit dimensions. An adaptively refined mesh with seven levels of refinement ($\approx 7.2 m$ data cells) was used.
Using our GPU-BH tree code and the CPU-BH tree code within FLASH4, we computed potentials with varying $\alpha$ values ($\alpha =$ 1.0, 0.75, 0.5, 0.25, 0.1). An example of the potential for $\alpha =0.25$ computed with the GPU-BH tree code using a maximum refinement level 7, is shown in fig.~\ref{LAURIN_FIG}. All simulations were run using 12 CPU-Cores with two GPU devices utilized by the GPU-BH tree code.\\\newline
We compare the analytical solution for the gravitational potential $\phi_\mathrm{MacLaurin}$ to the potential calculated with the GPU-BH tree code $\phi_\mathrm{GPU}$ and the CPU-BH tree code $\phi_\mathrm{CPU}$.
For the tests, we evaluate the relative error $\phi_\mathrm{err}$ with:
\begin{equation}
\phi_\mathrm{err}=\bigg | \frac{\phi_\mathrm{MacLaurin} - \phi_\mathrm{<algorithm>}}{\phi_\mathrm{MacLaurin}} \bigg |.
\label{ML_ERR}
\end{equation}
\\\newline
For comparison, we calculated the potential with the direct $O(n^2)$ method where a relative error of $2.17\times 10 ^{-4}$ was calculated. Typically, the Barnes \& Hut algorithm achieves higher accuracy with lower  $\alpha$ values \citep{Barnes:1986ed,1987ApJS...64..715H}. Figure~\ref{Errors} clearly shows that this is true for the GPU-BH tree code and for the CPU-BH tree code. Both, the GPU-BH tree code and the CPU-BH tree code produced relative errors below 1\%. For $\alpha$ values near $0.2$ both solvers nearly reached the accuracy of the direct summation $O(n^2)$ method. For $\alpha$ values between $0.25$ and $0.8$ the GPU-BH tree code calculated a more accurate potential than the CPU-tree code. For values above $0.8$ the CPU-BH tree achieved a higher accuracy. The differences between the GPU-BH and the CPU-BH accuracies are a direct result of the different strategies when applying the opening angle test.

\subsection{Scaling of the GPU-BH tree code with the top-hat sphere setup}
In this section, we summarize the results of different performance
tests based on the collapse of spherical cloud cores with and without
initial velocity fluctuations. The setup parameters of our
different simulations can be found in tables~\ref{tab_simulation_parameters} and~\ref{tab_simlist}.


\subsubsection{Simulation setup}\label{Parameters_Section}
We perform the scaling test with a top-hat (TH) sphere setup with the
following parameters (see also
table~\ref{tab_simulation_parameters}). We use a mean molecular weight
of $\mu=2.3\ \mathrm{g/mol}$ with an isothermal equation of state
(EoS) at a temperature of $20\ \mathrm{K}$. The density profile of the system with a sphere radius of $R_\mathrm{0}$ is described by a step function
\begin{equation}
\rho= \begin{cases}
  \langle  \rho \rangle & for\ r \leq R_\mathrm{0}\\
     0.01 \langle \rho \rangle & for\ r  > R_0.
\end{cases}
\label{topeq}
\end{equation}
With the density $\langle \rho \rangle =1.76 \times 10^{-18}\ \mathrm{g\ cm}^{-3}$ leading to a free fall time of
\begin{equation}
t_\mathrm{ff}=\sqrt{\frac{3\pi}{32G\langle\rho\rangle}}= 1.58 \times 10^{12} \mathrm{s} \approx 50\ \mathrm{kyr}.
\end{equation}
The initial sphere is highly gravitationally unstable with the Jeans length 
\begin{equation}
\lambda_\mathrm{J}=\sqrt{\frac{\pi c_\mathrm{s}^2}{G\langle\rho\rangle}}=5736\ \mathrm{au} = 0.28\ R_0.
\end{equation}
\newline
We use the top-hat sphere with varying parameters for our performance tests. To investigate the performance influence of varying GPU/CPU ratios, we use different sphere radii ($R_\mathrm{0}$) and refinement levels. One simulation setup with a sphere radius of $R_\mathrm{0}  = 3.5 \times 10^{17} \mathrm{cm}$ and $l_\mathrm{max} = 6$ resulting in $\approx 7 \times 10^6$ data cells (SP\text{-}RSS) and one with a radius of $R_\mathrm{0}  = 1.73\times 10^{17} \mathrm{cm}$ and $l_\mathrm{max} = 7$ resulting in $\approx 20 \times 10^6$ data cells (SP\text{-}RSL).\\
For the weak scaling test simulations (SP\text{-}WS), we scale the
problem size to the number of processing units by modifying the sphere
radius $R_0$. Here, we use maximum refinement level of
$l_\mathrm{max}= 8 \Rightarrow \Delta x = 7.8125\times 10^{14}\
\mathrm{cm}$. 
We also performed collapse simulations with the turbulent TH setups
(SP-TL) for a runtime of exactly one week. Here, we initialize the TH-profile
with an additional random supersonic velocity field with a Mach number
of $M_\mathrm{a} = 2.0$. The average turbulence crossing time is $t_\mathrm{tc}(R_0)=1.8 \times 10^5\ \mathrm{yrs}$  which is about 3.5 times larger than the free fall time $t_\mathrm{ff}$. An overview of the physical parameters is given in table~\ref{tab_simulation_parameters}.
In these simulations, we set the maximum refinement level to
$l_\mathrm{max}=15$ which results in a maximum resolution of 131072
grid cells in one direction, corresponding to $\approx \Delta x
\approx 0.4\ \mathrm{au}$. For the collapse simulations, we applied the Truelove criterion \citep{1997ApJ...489L.179T} to resolve the Jeans length \begin{equation}
\lambda_\mathrm{J}=\sqrt{\frac{\pi c_\mathrm{s}^2}{G\rho_\mathrm{max}}}
\end{equation}
throughout the simulation. Furthermore, we used the sink particle
module with the Federrath criterion for sink creation
\citep{Federrath:2010tu}. The sink particles have an accretion radius of $r_\mathrm{acc}=3\Delta x$ which leads to the threshold density $\rho_\mathrm{max}$ of
\begin{equation}
\rho_\mathrm{max}=\frac{\pi c_\mathrm{s}^2}{ 4G\,(3\,\Delta x)^2} = 2.518 \times 10^{-11}\ \mathrm{g\ cm}^{-3}
\end{equation}


For the strong scaling tests, we use the SP\text{-}TL simulation setup with a resolution of 12 grid cells per jeans length. We run the simulation with our accelerated gravity solver until two collapse regions including several sink particles are formed ($t_s=1.2 \times 10^{12}\ \mathrm{s} \approx 3.8 \times 10^4\ \mathrm{yrs}$). The corresponding checkpoint file is then used as an initial structure for the strong scaling test simulations (SP\text{-}TRS) with our accelerated solver, the CPU-BH tree solver and the Gridsolver. \\

\begin{table}[]%
\centering
\caption{Simulation parameters for the turbulent top-hat sphere}\renewcommand{\tabcolsep}{0.65pc}
\renewcommand{%
\arraystretch}{1.2}%
\small
\begin{tabular}{@{}llc}
\hline
\textbf{Parameter } &  & \textbf{Value} \\ \hline
Simulation box size  & $L_\mathrm{box}$  &   $8 \times 10^{17}\ \mathrm{cm}$ \\
Smallest cell size  & $\Delta x$ &  $6.1 \times 10^{12}\ \mathrm{cm}$ \\
Max. refinement & $l_\mathrm{max}$ &   15 \\
Min. refinement & $l_\mathrm{min}$ &    4\\
Sink particle accr. radius & $r_\mathrm{accr}$&    $1.83 \times 10^{13}\ \mathrm{cm}$ \\
Max. density & $\rho_\mathrm{max}$& $2.52 \times 10^{-11}\ \mathrm{g\ cm}^{-3}$\\
Opening angle parameter &$ \alpha$& 0.5 \\
Sphere radius & $R_0$  &   $3\times 10^{17}\ \mathrm{cm}$ \\
Total sphere mass & $M_\mathrm{tot}$ &  $ 100\ \mathrm{M}_{\odot}$ \\
Mean density & $\langle \rho \rangle$ &   $1.76 \times 10^{-18}\ \mathrm{g\ cm}^{-3}$ \\
Max. gas density & $\rho_\mathrm{max}$& $9.67\times 10^{-12}\ \mathrm{g\ cm}^{-3}$\\
Sound speed & $c_\mathrm{s}$ &    $0.27 \times 10^4\ \mathrm{km\ s}^{-1}$\\
Mean free fall time& $t_\mathrm{ff}$ & $1.58 \times 10^{12}\ \mathrm{s}\approx 50\ \mathrm{kyr}$ \\
Turbulent crossing time&$t_\mathrm{tc}$&$5.5\times 10^{12}\ \mathrm{s}$\\
Jeans Mass  &$M_\mathrm{J}$&$1.23\ \mathrm{M}_\sun$\\
Jeans length &$ \lambda_\mathrm{J}$&$1.4 \times 10^{17} \mathrm{cm}$ \\
\hline
\end{tabular}
\\[2pt]
\justify
\footnotesize{Physical and numerical simulation parameters for the turbulent top-hat sphere simulations.}
\label{tab_simulation_parameters}
\end{table}

\begin{table*}[]%
\centering
\caption{Scaling of the different top-hat sphere simulations}\renewcommand{\tabcolsep}{0.5pc}
\renewcommand{%
\arraystretch}{1.2}%
\small
\begin{tabular}{@{}lp{1.5cm}cp{2.0cm}clcp{2.0cm}p{2.0cm}}
\hline
\textbf{Name} &\textbf{Solver}&\textbf{Steps}&$\boldsymbol{R_\mathrm{0}[\mathrm{cm}]}$&$\boldsymbol{l_\mathrm{max}}$&$\boldsymbol{\Delta x\ [\mathrm{cm}]}$&\textbf{Resolution [cells]}&\textbf{Cell count} &\textbf{Figure}\\ \hline
\multirow{2}{*}{SP\text{-}RSS} &CPU-BH         & 10 &  $3.50\times 10^{17}$    & 6 &$3.12\times 10^{15}$& $256^3$&$7\times 10^6$&\ref{speedup-factors}\\
                                                           &GPU-BH         & 10 &  $3.50\times 10^{17}$    & 6 &$3.12\times 10^{15}$& $256^3$&$7\times 10^6$&\ref{speedup-factors}, \ref{Strong_1} \\
      SP\text{-}RSL                       &GPU-BH         & 10 &  $1.73\times 10^{17}$  & 7 &$1.56\times 10^{15}$& $512^3$ &$20\times 10^6$&\ref{TH_20}, \ref{speedup-factors}, \ref{Strong_1} , \ref{hy_perc}\\
\hline
\multirow{6}{*}{SP\text{-}WS}&GPU-BH& 10 &  $4.50\times 10^{16}\ \newline \text{-}\  \newline 3.60 \times 10^{17}$  & 8 &$7.81\times 10^{14}$&$1024^3$&$2.73\times 10^6\ \newline \text{-}\  \newline 1.48 \times 10^8$&\ref{WEAK_1}, \ref{WEAK_4}, \ref{WEAK_3}\\
                                                        &CPU-BH& 10 &  $4.50\times 10^{16}\ \newline \text{-}\  \newline 2.61 \times 10^{17}$  & 8 &$7.81\times 10^{14}$&$1024^3$&$2.73\times 10^6\ \newline \text{-}\  \newline 7.42 \times 10^7$&\ref{WEAK_1}\\
\hline
\multirow{2}{*}{SP\text{-}TL} &CPU-BH& 2006 &  $3.0\times 10^{17}$  & 15 &$6.10\times 10^{12}$&$131072^3$&$3.5\times 10^7$&\ref{TOPHAT_DENS}, \ref{PROP_LONG} \\
                                                       &GPU-BH& 4465 &  $3.0\times 10^{17}$  & 15 &$6.10\times 10^{12}$&$131072^3$&$4.4\times 10^7$&\ref{TOPHAT_DENS}, \ref{PROP_LONG}, \ref{contrib_long_routines}, \ref{contrib_long_routines} \\
\hline
\multirow{5}{*}{SP\text{-}TS} &GPU-BH & 2 &  $3.0\times 10^{17}$  & 15 &$6.10\times 10^{12}$&$131072^3$ &$1.85\times 10^7$& \ref{turb_zoom1}, \ref{turb_zoom2}, \ref{strong_turb}, \ref{strong_turb_solvers}, \ref{strong_speed}, \ref{strong_turb_data}, \ref{turb_ratio}\\
                                                           &CPU-BH & 2 &  $3.0\times 10^{17}$  & 15 &$6.10\times 10^{12}$ &$131072^3$&$1.85\times 10^7$& \ref{turb_zoom1}, \ref{turb_zoom2}, \ref{strong_turb_solvers}, \ref{strong_speed}, \ref{strong_turb_data}, \ref{turb_ratio}\\
                                                           & Gridsolver & 2 &  $3.0\times 10^{17}$  & 15 &$6.10\times 10^{12}$&$131072^3$ &$1.85\times 10^7$& \ref{turb_zoom1}, \ref{turb_zoom2}, \ref{strong_turb_solvers}, \ref{strong_speed}, \ref{strong_turb_data}, \ref{turb_ratio}\\
\hline
\end{tabular}
\\[2pt]
\justify
\footnotesize{Top-hat sphere simulations used for our scaling tests. The cell count for the SP\text{-}TL setups refer to the final stage of the simulations. The radii for the the SP\text{-}WS simulations are used to scale the cell count for the weak scaling test simulations (see table~\ref{tab_blockcount} for a list of radii and their corresponding cell count).}
\label{tab_simlist}
\end{table*}

\subsubsection{Scaling with different GPU/CPU ratios}\label{Scaling with different GPU/CPU ratios section}
\begin{figure}
 \includegraphics[width=0.5\textwidth]{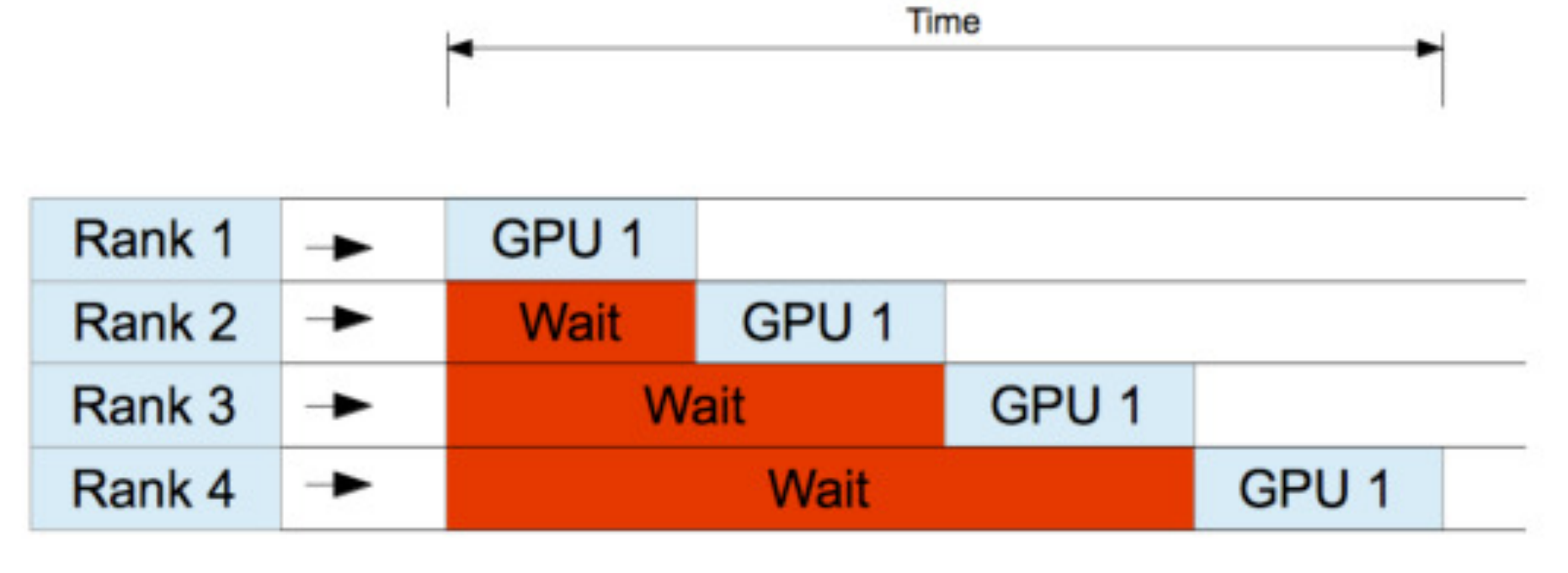}%
\caption{Illustration of the serial access sequence for processes competing for the same GPU device. Direct access to the GPU device is only granted for the first process in the queue (Rank 1). All other processes wait for the GPU device to be available again.}
\label{compete}
\end{figure}

\begin{figure}
 \includegraphics[width=0.5\textwidth]{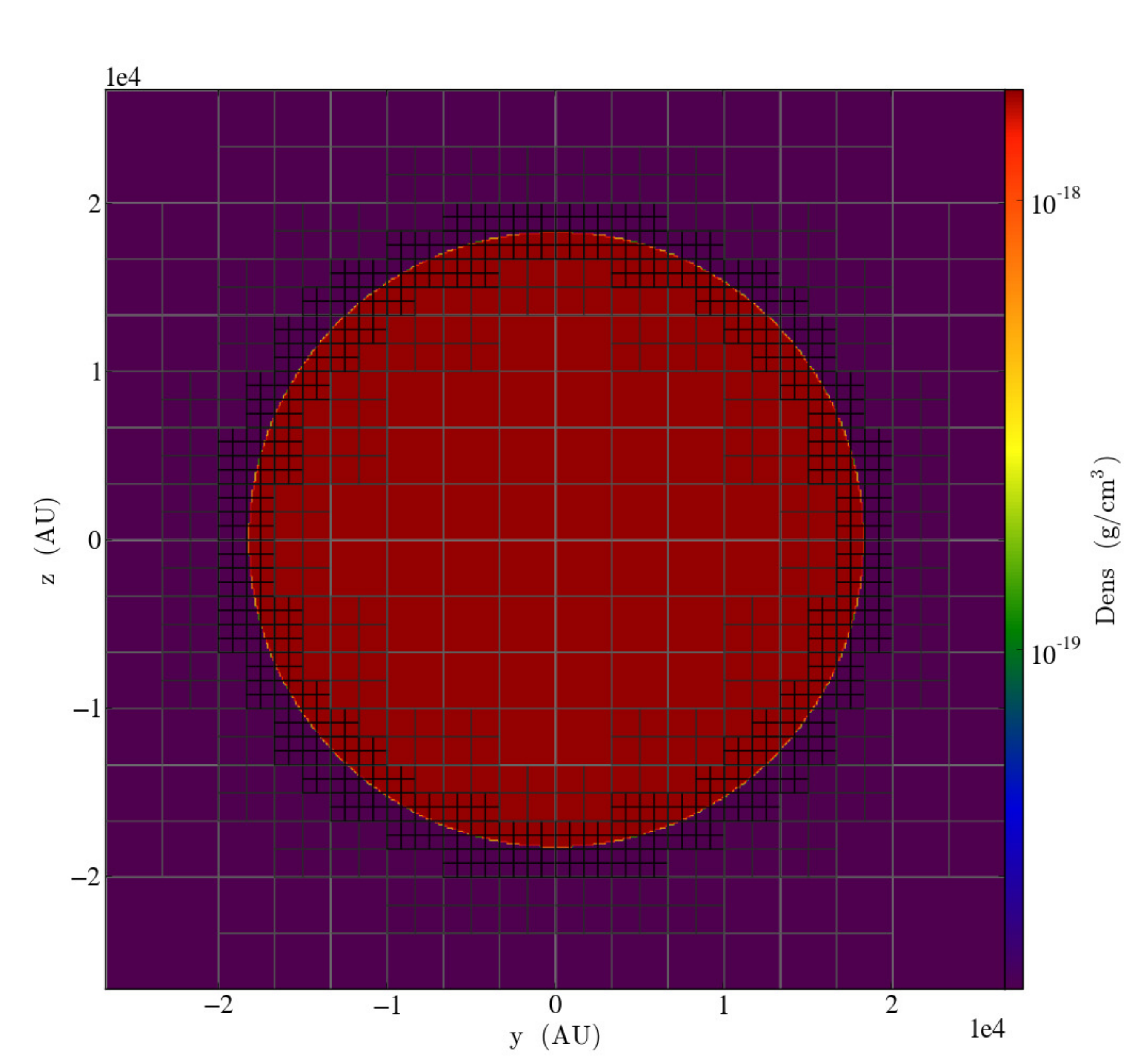}%
\caption{The initial density and AMR block distribution for the SP\text{-}RSL setup. The image shows a slice through the box along the X-axis. An adaptively refined mesh with seven levels of refinement was used with a sphere radius of $1.73\times10^{17}$ resulting in $\approx 20 \times 10^6$ data cells.}
\label{TH_20}
\end{figure}

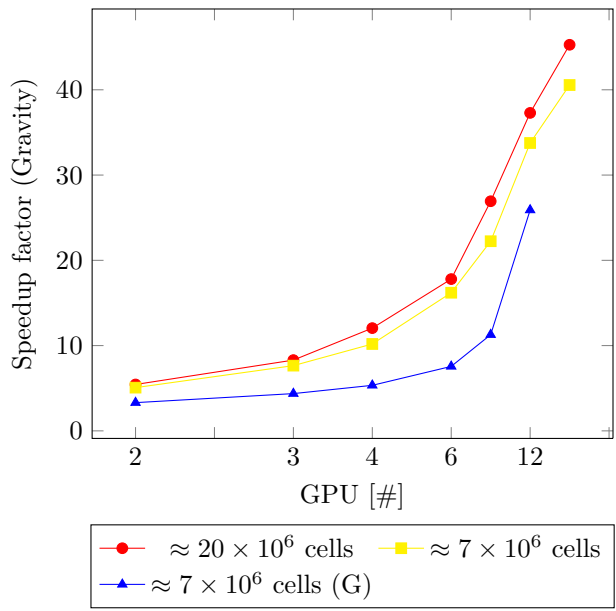
\begin{figure}
\begin{tikzpicture}
 \begin{axis}[
   xticklabels={ ,12,6,4,3, ,2},
    xtick={0,2,4,6,8,10,12},
    xlabel={GPU [\#]},
    ylabel={Speedup factor (Gravity)},
    scaled x ticks=false,
    x dir=reverse,
      legend style={at={(0.5,-0.2)},
    anchor=north,legend columns=2},
    legend entries={$\approx 20 \times 10^6$ cells ,$\approx 7 \times 10^6$ cells,$\approx 7 \times 10^6$ cells (G)}
]
\addplot [
color=red, 
mark = *] 
table [
x = CPU,
y = SPUP
   ] {
     SPUP CPU
      5.44   12
      8.30   8  
    12.05   6 
    17.80   4 
    26.93   3
    37.29   2
    45.28   1
};
\label{red_ratio}
\addplot [
color=yellow,
mark = square*
]
table[
x = CPU,
y = SPUP 
] {
     SPUP CPU
      5.07   12  
      7.65   8 
    10.19   6 
    16.20   4 
    22.24   3
    33.76   2
    40.56   1
};
\label{yellow_ratio}
\addplot [
color=blue,
mark =triangle*] 
table [
x = CPU,
y = SPUP
   ] {
     SPUP CPU
      3.31   12
      4.37   8  
      5.34   6 
      7.56   4 
    11.28   3
    25.90   2
};
\label{blue_ratio}
\end{axis}
\end{tikzpicture}
\caption{Speedup factors for the gravity module using different GPU/CPU ratios for the GPU-BH tree code in comparison to the CPU-BH tree code. The Y-axis refers to the speedup of the GPU-BH tree code in comparison to the CPU-BH tree code and the X-axis refers to the GPU count for different GPU/CPU ratios using a total of 24 CPU cores. We see the highest speedup with the  SP\text{-}RSL setup (\ref{red_ratio}). With the smaller simulation setup SP\text{-}RSS (\ref{yellow_ratio}), the speedup is slightly lower especially when using less CPU cores per GPU. The speedup factors with the SP\text{-}RSS setup-using the gathered data approach (\ref{blue_ratio}) show the lowest speedup compared to the other setups. The highest speedup of $\approx 45$ was gained for a 1/1 GPU/CPU ratio (24 GPU devices and 24 CPU cores) with the SP\text{-}RSL simulation setup holding $\approx 20 \times 10^6$ data cells (\ref{red_ratio}).}
\label{speedup-factors}
\end{figure}

In general, FLASH 4 \citep{2000ApJS..131..273F} simulations are setup for distributed-memory machines where several MPI \citep{Gabriel:2004ub} processes are created. Since every process holds its own CUDA context and kernels are effectively serialized as they launch on the GPU, several MPI ranks compete for the device and slow down the code execution. Figure~\ref{compete} shows an example sequence of the access order for 4 processes trying to access the same GPU device (4 MPI Ranks per compute node).\\
In order to compare the runtimes of the GPU-BH tree code at different CPU/GPU ratios with the runtime of the CPU-BH code, we calculate the respective speedup $S$ with:
\begin{equation}
\label{speedup_eq}
S=\frac{T_\mathrm{CPU-BH}}{ T_\mathrm{GPU-BH}}
\end{equation}
where $T_\mathrm{CPU-BH}$ is the time for one gravity step evaluated
for the CPU-BH tree code and $T_\mathrm{GPU-BH}$ is the time for one
gravity step evaluated for the GPU-BH tree code. Independent from the
problem size (number of data cells), the highest speedup is expected
for a one to one GPU/CPU ratio.
We analyzed the performance depending on different GPU/CPU ratios with
the SP\text{-}RSS run using $\approx 7 \times 10^6$ cells and
with the SP\text{-}RSL run with $\approx 20 \times 10^6$ cells (see fig.~\ref{TH_20}).\\
Figure~\ref{speedup-factors} shows the speedup calculated with
Eq.~\ref{speedup_eq} for GPU/CPU ratios of  1/12, 1/8, 1/6, 1/4, 1/2
and 1/1 while maintaining a constant number of 24 cores for each
ratio. Here, we reach a maximum speedup of $\approx 45$ with the one-to-one GPU/CPU ratio (24 GPU devices and 24 CPU cores) for the simulation with $\approx 20 \times 10^6$ data cells (fig.~\ref{speedup-factors},~\ref{red_ratio}). Similarly, the lowest speedup factor of $\approx 5$ was evaluated for the 1/12 ratio (2 GPU devices and 24 CPU cores) caused by the long GPU device access times following the serialized access pattern illustrated in fig.~\ref{compete}.

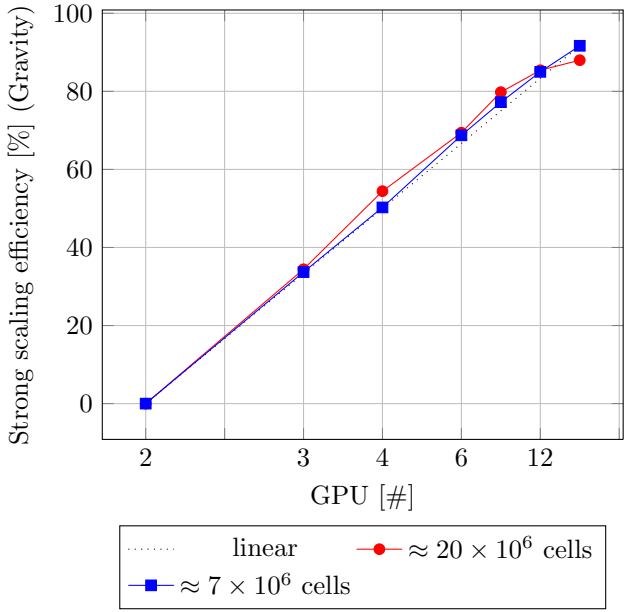
\begin{figure}
\begin{tikzpicture}
 \begin{axis}[
    xticklabels={ ,12,6,4,3, ,2},
    xtick={0,2,4,6,8,10,12},
    xlabel={GPU [\#]},
    ylabel={Strong scaling efficiency [\%] (Gravity)},
    grid=major,
    x dir=reverse,
    legend style={at={(0.5,-0.2)},
    anchor=north,legend columns=2},
    legend entries={linear ,$\approx 20 \times 10^6$ cells,$\approx 7 \times 10^6$ cells}
    ]

\addplot [
smooth,
color=black, 
dotted,] 
table [
x = CPU,
y = SPUP
   ] {
     SPUP CPU
     0             12  
    33.33      8  
    50            6 
    66.66      4 
    75            3
    83.33      2
    91.67      1
};
\label{strong_ratio_ideal}

\addplot [
color=red, 
mark =*] 
table [
x = CPU,
y = SPUP
   ] {
     SPUP CPU
        0         12  
    34.42     8  
    54.42     6 
    69.41     4 
    79.79     3
    85.40     2
    87.98     1
};
\label{strong_ratio_large}
\addplot [
color=blue,  
mark = square*
]
table[
x = CPU,
y = SPUP 
] {
     SPUP CPU
     0.00    12
      33.7   8  
      50.26 6  
    68.73   4 
    77.22   3 
    84.99   2
    91.667 1
};
\label{strong_ratio_small}
\end{axis}
\end{tikzpicture}
\caption{The strong scaling efficiency of our GPU-BH tree solver calculated with Eq.~\ref{eq_sc} for varying CPU/GPU ratios and a constant number of CPU cores. The X-axis refers to the GPU count for different GPU/CPU ratios using a total of 24 CPU cores. We see a nearly linear strong scaling for the SP\text{-}RSS simulation (\ref{strong_ratio_small}) and the SP\text{-}RSL simulation (\ref{strong_ratio_large}).}
\label{Strong_1}
\end{figure}

One way to circumvent this performance bottleneck is to collect all data on one of the CPU cores sharing one device (fig.~\ref{speedup-factors},~\ref{blue_ratio}).\footnote{This is only practical for small problem sizes because the GPU-device memory is comparably small and limited.}  Although intra node communication is comparably fast and the number of required tree builds is reduced, the evaluated speedup factors for this approach are generally lower compared to the previously described approach.\\
\noindent
Nevertheless, we achieved a nearly linear scaling efficiency
considering the count of GPU-devices. In Fig.~\ref{Strong_1} we show the strong scaling efficiency for different GPU to CPU ratios. Starting with a time $t_\mathrm{1}$ for one gravity step at a GPU/CPU ratio of 1/12, the percentual efficiency at a specific ratio with the respective time $t_\mathrm{r}$ is given as:
\begin{equation}
\label{eq_sc}
SC_\mathrm{ratio}= \frac{t_\mathrm{1} - t_\mathrm{r}}{t_\mathrm{1}} *100
\end{equation}
The results are shown for the two simulations SP-RSS and SP-RSL.
Note, that the base ratio of 1 GPU device for 12 CPU cores  ($t_\mathrm{1}$) is achieved using 2 compute nodes with a total of 24 cores and 2 GPU devices and already includes communication times for inter and intra node communication. (The presented speedup and scaling values refer to the gravity solver module only.) Although we measured a speedup factor of $\approx 40$ compared to the original CPU-BH tree code (see fig.~\ref{speedup-factors},~\ref{red_ratio}), the entire wall time for the simulation did not show the same speedup since our gravity solver only makes up less than half of the entire wall time for the related GPU/CPU ratio. \\

\begin{figure}
\begin{tikzpicture}
	\begin{axis}[
          xlabel={GPU [\#]},
          ylabel={Runtime fraction [ \%]},
          grid=major,
	ybar stacked,
	ymin=0,
	symbolic x coords={2, 3, 4, 6, 8, 12, 24,0},
	legend style={at={(0.5,-0.2)},
         anchor=north,legend columns=-1}
		]

	\addplot[green,
	               fill=green!50] coordinates
		{(2,65) (3,56) (4,45) (6,39) (8,31) (12,26) (24,23)};
		\label{grv_bar}
          \addplot[blue,
                    fill=blue!50] coordinates
		{(2,35) (3,44) (4,55) (6,61) (8,69) (12,74) (24,77)}; 
		\label{hy_bar}
         \legend{Gravity, Hydro}
	 \end{axis}
\end{tikzpicture}
\caption{Runtime fractions of the ppm hydro solver (~\ref{hy_bar} ) and our accelerated gravity solver (~\ref{grv_bar} ) for different CPU/GPU ratios. The X-axis refers to the GPU count for different GPU/CPU ratios using a total of 24 CPU cores. Values are evaluated for the SP\text{-}RSL simulation with $\approx 20 \times 10^6$ data cells. With a lower runtime of the gravity solver for increasing GPU/CPU ratios, the hydro solver becomes the dominating solver for the simulations. For a CPU/GPU ratio of 6/1 (4 GPU devices and 24 CPU cores), already $55\%$ of the evolution time was spent in the hydro routines.}
\label{hy_perc}
\end{figure}
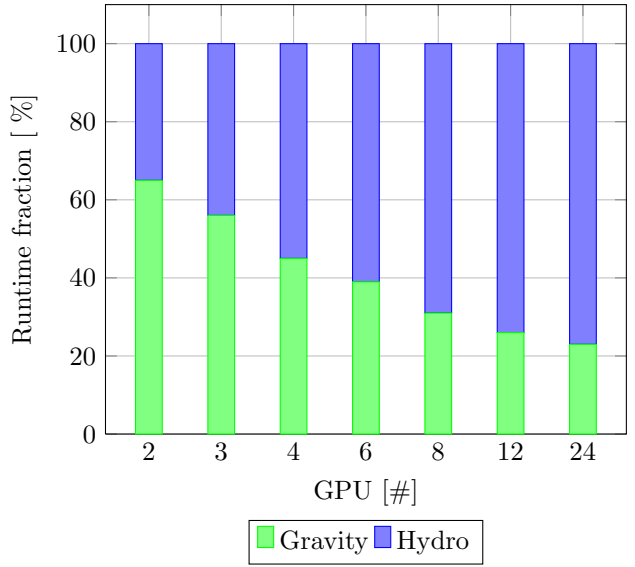

Figure~\ref{hy_perc} shows the percentual fractions of our GPU accelerated gravity solver and the hydro solver for different GPU/CPU ratios. The presented values are evaluated for the (SP\text{-}RSL) simulation with $\approx 20 \times 10^6$ data cells.\footnote{Values evaluated for the smaller SP\text{-}RSS simulation with $\approx 7 \times 10^6$ data cells only differ for $\approx 2\%$ to the presented values.}
At a ratio of 1/12, we find the runtime share of our gravity solver is the dominating part with $\approx 65\%$ of the total wall time.\footnote{For the same simulation setup the CPU-BH gravity solver dominated the runtime with $\approx  88\%$.} 
Since our accelerated gravity solver benefits from a lower GPU/CPU ratio, its contribution to the entire simulation time is reduced with lower ratios. Hence, we measure lower runtime shares with lower ratios. Already for a ratio of 1/6, our solver contributes only $\approx 45 \%$ to the wall time. Finally, we measure only a $\approx 23 \%$ contribution for a ratio of 1/1 and find the hydro solver to be the dominating code part (see fig.~\ref{hy_perc}).


\subsubsection{Weak scaling with the top-hat sphere}\label{Weak scaling with the top-hat sphere section}
\begin{figure}[]
\begin{tikzpicture}
 \begin{axis}[
    xlabel={$P_\mathrm{U}$ [\#]},
    ylabel={Parallel efficiency  [\% of  linear scaling] (Gravity)},
         legend style={at={(0.5,-0.2)},
         anchor=north,legend columns=1},
    legend entries={$\approx 8.68 \times 10^5$ cells/cores,$\approx 6.72 \times 10^5$  cells/cores,$\approx 4.4 \times 10^5$  cells/cores,$\approx 4.4 \times 10^5$  cells/cores (CPU)},
    ]
\addplot [
color=red, 
mark =*] 
table [
x = CPU,
y = SPUP
   ] {
       SPUP CPU
  100.0    1
   95.9    4  
   93.3    8 
    91.5  12 
    89.9  16
    90.4   20
    86.4   24
    88.7   28
};
\label{weak_2k}
\addplot [
color=green,  
mark = triangle*
]
table[
x = CPU,
y = SPUP 
] {
     SPUP CPU
     100.0    1
       98.5    4  
       92.6   8 
       89.1  12 
       88.2  16
       86.1   20
       84.3   24
       80.1   28
};
\label{weak_1.5k}
\addplot [
color=blue,  
mark = square*
]
table[
x = CPU,
y = SPUP 
] {
     SPUP CPU
     100.0    1
       98.9   4  
       90.5    8 
       88.3  12 
       83.9  16
       81.0   20
       80.9   24
       78.8   28
};
\label{weak_1k}
\addplot [
color=black,  
mark = *
]
table[
x = CPU,
y = SPUP 
] {
     SPUP CPU
     100.0    1
       83.0    4  
       86.8    8 
       80.0  12 
       82.5  16
       82.9  20
       82.6  24
       80.7  28
};
\label{weak_cpu}
\end{axis}
\end{tikzpicture}
\caption{Weak scaling efficiency in \% of linear for the top-hat simulations with the SP\text{-}WS setup (gravity only). The scaling values for the smaller simulations with $4.4 \times 10^5$ data cells/cores run with our accelerated solver (\ref{weak_1k})  and with the CPU-BH solver~\ref{weak_cpu} show a similar scaling trend for more then 15 $P_\mathrm{U}$. We achieved the best scaling results for our solver with the larger system holding $\approx 8.68 \times 10^5$ data cells/cores (\ref{weak_2k}).}
\label{WEAK_1}
\end{figure}
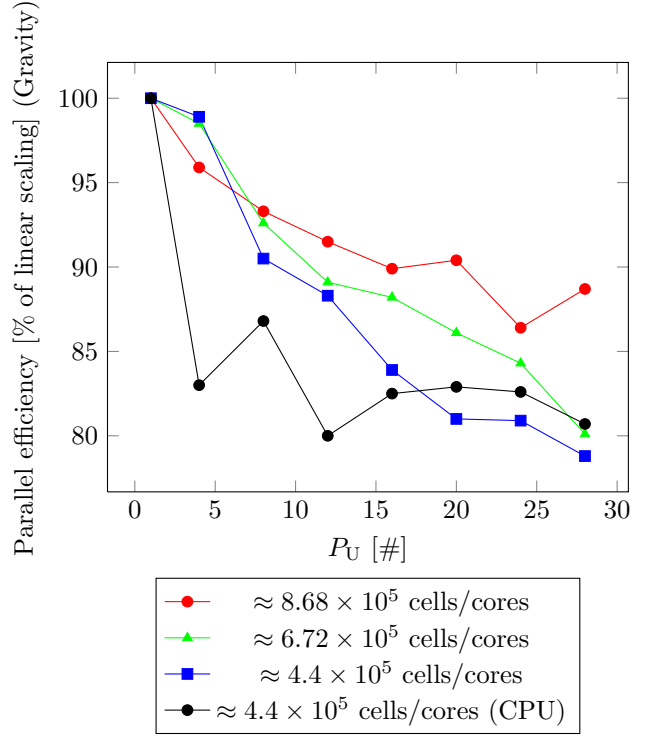

\begin{figure}[]
\begin{tikzpicture}
	\begin{axis}[
          xlabel={$P_\mathrm{U}$ [\#]},
          ylabel={Runtime fraction [\%]},
          grid=major,
	ybar stacked,
	ymin=0,
	symbolic x coords={1,4,8,12,16,20,24,28},
	legend style={at={(0.5,-0.2)},
         anchor=north,legend columns=-1},
		]
		
   \addplot
coordinates
                  {   
(1, 0.720)
(4, 1.705)	
(8, 3.008)	
(12, 4.247)	
(16, 5.442)	
(20,5.442 )	
(24, 7.749)	
(28, 9.329)
};

\addplot
	               coordinates
		{
		(1, 97.158)	
		(4, 94.662)	
		(8, 92.348)
		(12, 90.488)	
		(16, 88.865)	
		(20, 88.865)			
		(24, 85.240)	
		(28, 83.279)
		};

	\addplot
	               coordinates
		{ 
		 (1, 1.064)
		(4,  1.950)	
		(8,  2.673)	
		(12, 3.218)	
		(16, 3.713)	
		(20, 3.713)	
		(24, 4.617)	
		(28, 5.047)
		};        		
				
		\addplot
	               coordinates
		{
		(1, 1.058)	
		(4, 1.683)	
		(8, 1.970)
		(12, 2.048)	
		(16, 1.980)	
		(20, 1.980)			
		(24, 2.384)	
		(28, 2.345)
		}; 	
         \legend{E. kernel,  F. kernel, E. memory, F. memory}
	 \end{axis}
\end{tikzpicture}
\caption{Weak scaling of the runtime fractions of GPU device memory operations and GPU kernel executions. The blue and the yellow bars refer to the runtimes of the memory operations and the kernel executions during the calculation of the Essential Nodes. The red and the grey bars show the proportions for the code calculating the final potential. The values are calculated for the SP\text{-}WS simulation with $\approx 8.68 \times 10^5$ data cells. One $P_\mathrm{U}$ refers to 6 CPU cores and 1 GPU device.}
\label{WEAK_4}
\end{figure}
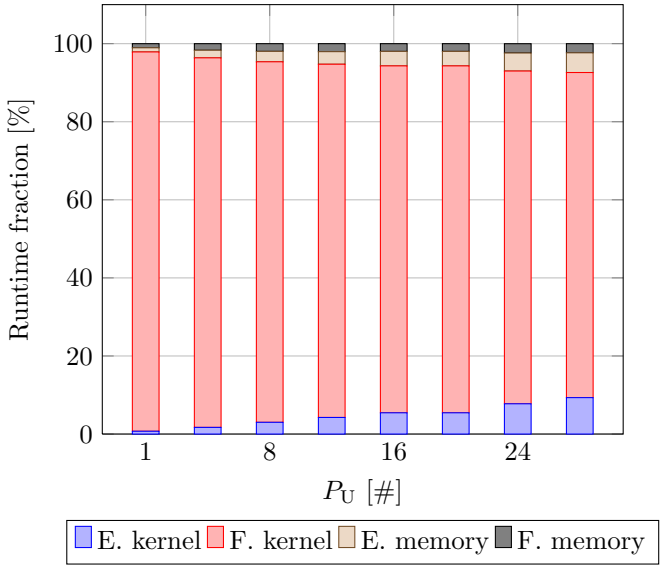

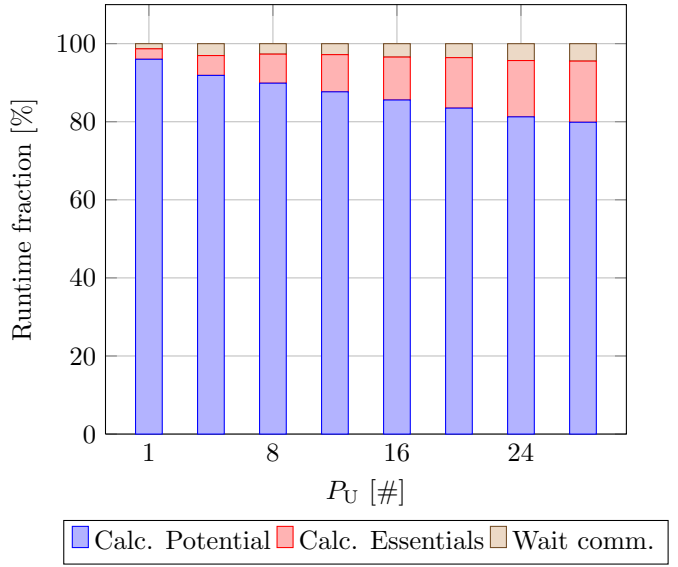
\begin{figure}[]
\begin{tikzpicture}
	\begin{axis}[
          xlabel={ $P_\mathrm{U}$ [\#] },
          ylabel={Runtime fraction [\%]},
          grid=major,
	ybar stacked,
	ymin=0,
	symbolic x coords={1,4,8,12,16,20,24,28},
	legend style={at={(0.5,-0.2)},
         anchor=north,legend columns=-1},
		]
		
   \addplot
coordinates
                  {   
(1, 96.0033062673214)
(4, 91.8707403055229)	
(8, 89.8750644864635)	
(12, 87.6642917201398)	
(16, 85.5670328708364)	
(20, 83.4905559477208)	
(24, 81.2515953373607)	
(28, 79.8698463521633)
};
		
	\addplot
	               coordinates
		{ 
		 (1, 2.6790489619293)
		(4,  5.0834312573443)	
		(8, 7.44454164143283)	
		(12, 9.523388060362)	
		(16, 10.9970674486804)	
		(20, 12.9307578771738)	
		(24, 14.4261039734536)	
		(28, 15.6641786390004)
		};        		
		\addplot
	               coordinates
		{
		(1, 1.31764477074926)	
		(4, 3.04582843713279)	
		(8, 2.67815086467936)
		(12, 2.81010753639864)	
		(16, 3.43371120934915)	
		(20, 3.57440800872746)			
		(24, 4.31804645622394)	
		(28, 4.46181674532715)
		};

         \legend{Calc. Potential, Calc. Essentials, Wait comm.}
	 \end{axis}
\end{tikzpicture}
\caption{Weak scaling runtime fractions of different parts of the GPU-BH tree gravity solver for the SP\text{-}WS simulation with $\approx 8.68 \times 10^5$ data cells per CPU core. The runtime fraction of the routines for calculating the Essential Nodes grows with increasing number of $P_\mathrm{U}$. One $P_\mathrm{U}$ refers to 6 CPU cores and 1 GPU device.}
\label{WEAK_3}
\end{figure}

For the weak scaling, the problem size (number of data cells) assigned
to each processing unit stays constant. We refer to a processing unit,
``$\mathrm{P_U}$'' as a unit of six CPU cores plus one GPU
device.\footnote{We choose six cores to fully load exactly one of the hexacore CPUs on each compute node and use the MPI call to control the respective processor binding affinity.} 
We use the weak scaling efficiency given by:
\begin{equation}
\label{WEAK_EQ}
\frac{t_\mathrm{1}}{(t_\mathrm{N})}*100 \, ,
\end{equation}
where $t_\mathrm{1}$ is the wall time of one gravity time step with
one $P_\mathrm{U}$ and $t_\mathrm{N}$ is the time of one gravity step
with N processing elements. We use the SP-WS top-hat sphere setup and scale the problem size to the number of $P_\mathrm{U}$ by changing the sphere radius $R_\mathrm{0}$ and as a result, the overall cell count. Table~\ref{tab_blockcount} gives an overview of the different sphere radii and cell counts used for the weak scaling tests. (The scaling results are normalized to match 1 Leaf-node per core.)\\
\begin{table}[]%
\caption{AMR block count with different sphere radii}\renewcommand{\tabcolsep}{0.5pc}
\renewcommand{%
\arraystretch}{1.2}%
\small
\begin{tabular}{@{}lcccl}
\hline
$R_\mathrm{0}$&Blocks&Leaf-blocks/Core&cells&$P_\mathrm{U}$ \\ \hline

\hline
\multicolumn{4}{c}{$SP−WS$ with $\approx 4.4\times 10^5$ data cells/cores}\\
\hline
 $4.50 \times 10^{16}$  &    6089      & $\approx 888$& $2.73\times 10^6$&1\\
 $9.70 \times 10^{16}$  &    24457   & $\approx 891$&  $1.10\times 10^7$ &4\\
 $1.36 \times 10^{17}$  &    44937    & $\approx 819$& $2.01\times 10^7$&8\\
 $1.70 \times 10^{17}$  &   71945     &$\approx 874$&$3.22\times 10^7$&12\\
 $1.98 \times 10^{17}$  &   94985     &$\approx 865$&$4.25\times 10^7$&16\\
 $2.20 \times 10^{17}$   & 118537    &$\approx 864$&$5.31\times 10^7$&20\\
 $2.40 \times 10^{17}$   & 141257    &$\approx 858$&$6.32\times 10^7$&24\\
 $2.61 \times 10^{17}$   & 165705 &$\approx 863$&$7.42\times 10^7$&28\\
\hline
\multicolumn{4}{c}{$SP−WS$ with $\approx 6.72\times 10^5$ data cells/cores}\\
\hline
 $5.70 \times 10^{16}$  &   8905 &$\approx 1298$&$3.99\times 10 ^6$&1\\
 $1.19 \times 10^{17}$ &  36617&$\approx 1335$&$1.64\times 10^7$&4\\
 $1.70 \times 10^{17}$  &  71945 &$\approx 1311$&$3.22\times 10^7$&8\\
 $2.08 \times 10^{17}$  & 105993 &$\approx 1288$&$4.75\times 10^7$&12\\
 $2.43 \times 10^{17}$   & 144905 &$\approx 1320$&$6.49\times 10^7$&16\\
 $2.70 \times 10^{17}$   & 178889 &$\approx 1320$&$8.01\times 10^7$&20\\
 $2.98 \times 10^{17}$   & 217033 &$\approx 1318$&$9.72\times 10^7$&24\\
 $3.23 \times 10^{17}$  &  253577 &$\approx 1320$&$1.14\times 10^8$&28\\
\hline
\multicolumn{4}{c}{$SP−WS$ with $\approx 8.68\times 10^5$ data cells/cores}\\
\hline
 $6.50 \times 10^{16}$ &  11721& $\approx 1709 $&$ 5.25\times 10^6$&1\\
 $1.36 \times 10^{17}$  &  44937 & $\approx 1638$&$2.01\times 10^7$&4\\
 $1.98 \times 10^{17}$  &  94985 &$\approx 1731$&$4.25\times 10^7$&8\\
 $2.40 \times 10^{17}$   & 141257 &$\approx 1716$&$6.32\times 10^7$&12\\
 $2.75 \times 10^{17}$   & 183497 &$\approx 1672$&$8.22\times 10^7$&16\\
 $3.10 \times 10^{17}$   & 232905 &$\approx 1698$&$5.25\times 10^6$&20\\
 $3.40 \times 10^{17}$   & 277193 &$\approx 1684$&$1.24\times 10^8$&24\\
 $3.60 \times 10^{17}$   & 330377 &$\approx 1720$&$1.48\times 10^8$&28\\
\hline
\end{tabular}
\\[2pt]
\footnotesize{The AMR block count for different sphere radii used to scale the top-hat sphere simulations $SP−WS_\mathrm{CPU}$ and $SP−WS_\mathrm{GPU}$. }
\label{tab_blockcount}
\end{table}
In fig.~\ref{WEAK_1} we show the weak scaling efficiency
(Eq.~\ref{WEAK_EQ}) for three different simulation sizes with the
SP\text{-}WS setup holding $\approx 8.68 \times 10^5$
(\ref{weak_2k}), $6.72 \times 10^5$ (\ref{weak_1.5k}) and $4.4 \times
10^5$ (\ref{weak_1k}) data cells assigned to each CPU core. We see a
drop of the weak scaling efficiency with increasing number of
processing elements. The main reason for this loss in efficiency is
not only based on increased communication times also the algorithm
design is a factor. Especially the method of calculating the essential
nodes plays a role, since the local tree is traversed once for every
process. Although this is beneficial to overlap computation with
communication time, the loop size increases with higher numbers of
processing elements. As a result, more data is copied from  device
memory to host memory and more GPU kernels are started.
The fractional runtimes of the respective memory operations and kernel
executions are outlined in fig.~\ref{WEAK_4}. With 16
$P_\mathrm{U}$, the memory operations and kernel executions during the
calculation of the Essential Nodes make up $\approx 10 \%$ of the
total GPU time which is dominated by kernel executions. Note, that the
actual runtime per kernel does not change but the kernel is started
more often.
The large number of kernel calls and memory operations is clearly
reflected in the runtime of the the routines for calculating the
Essential Nodes. Figure~\ref{WEAK_3} illustrates the growing runtime
contribution of the respective routines with the larger top-hat setup
holding $\approx 1 \times 10^5$ data cells per core. We see that the
routines for calculating the Essential Nodes take a maximum of
$\approx 15\%$ of the total solver runtime with 28 $P_\mathrm{U}$.
Nevertheless, at the same count nearly $\approx 80\%$ of the linear
scaling efficiency was reached with the small system and $\approx
88\%$ was reached with the larger system.
For the smaller setup, the CPU-BH  and the GPU-BH simulations show
nearly the same weak scaling efficiency for $\mathrm{P_U} \ge 15$. 

\subsubsection{Comparison of GPU-BH and CPU-BH for fixed runtime (7 days)}\label{Comparison of GPU-BH and CPU-BH for fixed runtime (7 days) section}

We run two simulations of the top-hat sphere setup with turbulence and
sink particles (SP\text{-}TL). One simulation with our GPU
accelerated gravity solver and one with the CPU-BH Tree solver. While
the CPU only simulation was run with $72$ CPU cores, we used $60$ CPU
cores and $10$ GPU-devices for the GPU accelerated simulation. Both
simulations were run for $t_\mathrm{w}=168\ \mathrm{h}=7\
\mathrm{days}$. With the CPU-BH simulation, we reached a total of
$2006$ evolution steps and a final simulation time of $t_\mathrm{s}
\approx 1.0 \times 10^4\ \mathrm{yrs}$. Since the simulation stopped
at an early stage of the collapse with a maximum density of
$9.12\times 10^{17}\ \mathrm{g}\ \mathrm{cm}^{-3}$, neither the
maximum refinement level was reached nor any sink particles were
formed.

\begin{figure*}[]
\subfigure[GPU-accelerated at $t_\mathrm{sim}=
  1.02 \times 10^4\ \mathrm{yrs}$, $t_\mathrm{wall}=62\
  \mathrm{h}$]{\includegraphics[width=0.5\textwidth]{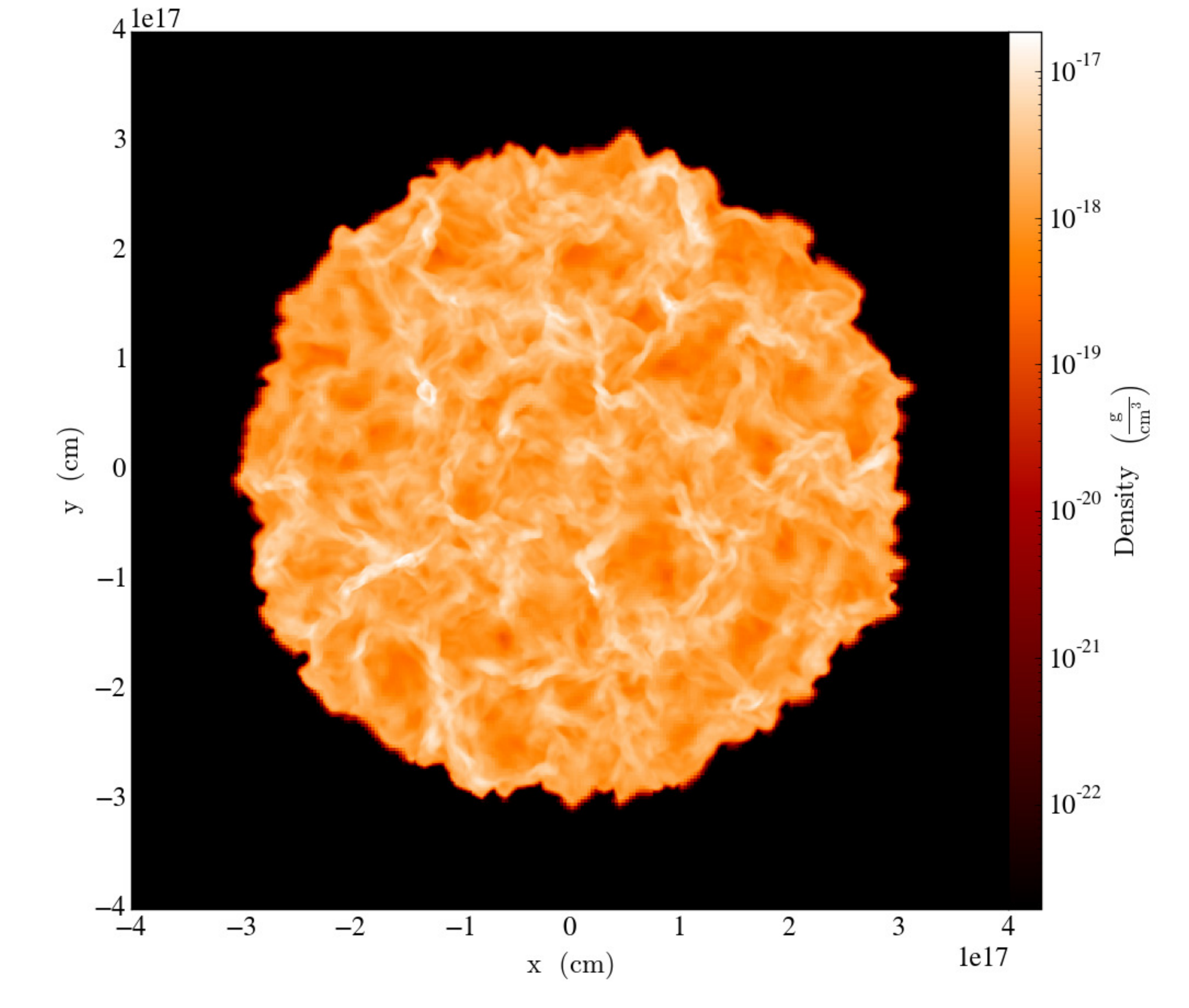}}
\subfigure[CPU
  only at $t_\mathrm{sim} = 1.02 \times 10^4\ \mathrm{yrs}$,
  $t_\mathrm{wall}=168\
  \mathrm{h}$]{\includegraphics[width=0.5\textwidth]{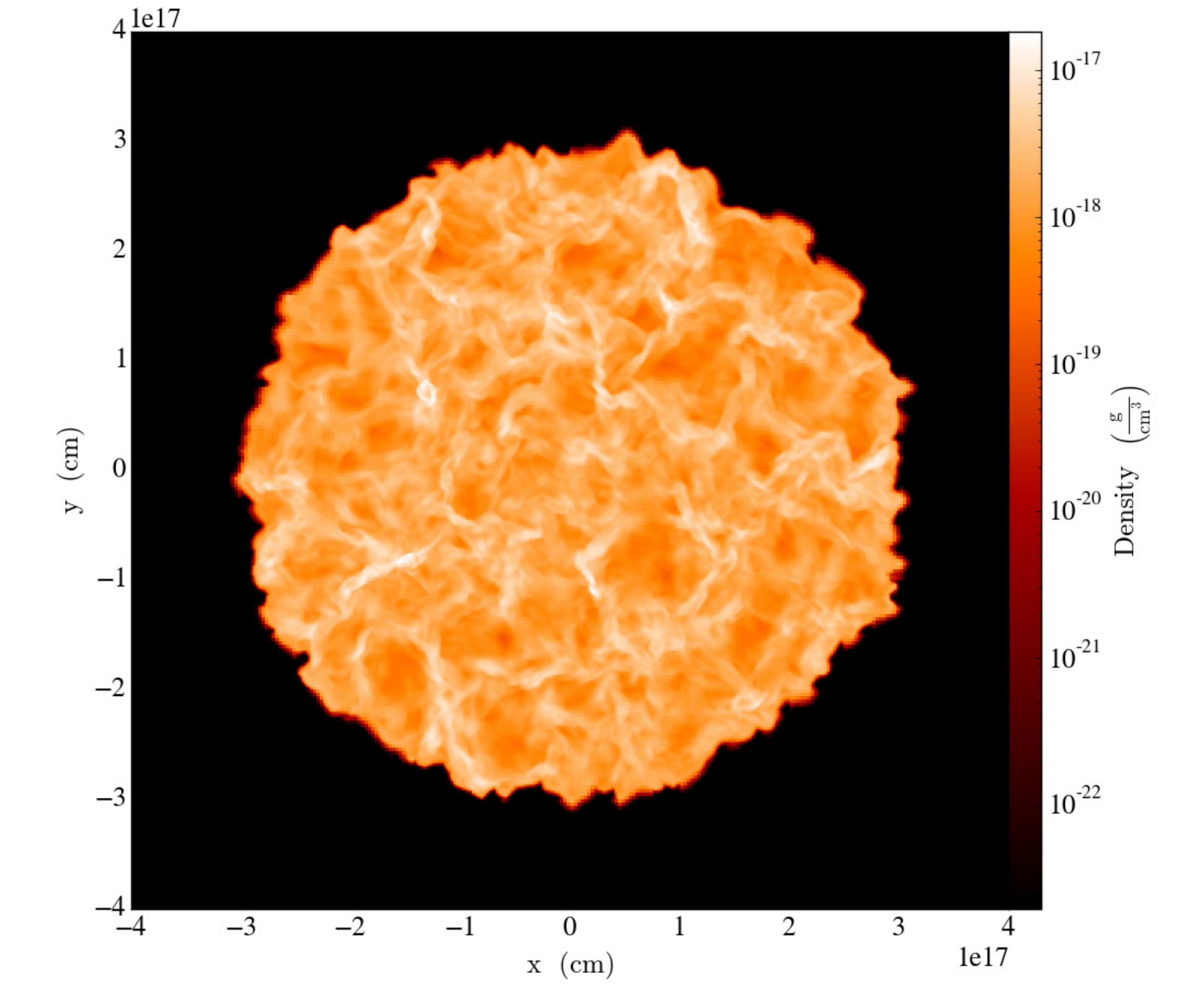}}
\caption{Slices
    of the density for  the SP-TL simulation at $t_\mathrm{sim} = 1.02 \times 10^4\ \mathrm{yrs}$. The left plot shows the GPU accelerated simulation while the right plot shows the CPU only simulation. The respective simulation step was reached after 7 days with the CPU only solver and after $\approx 2.6\ \mathrm{days}$ with the GPU accelerated solver. The CPU simulation was executed with 72 CPU cores and the accelerated GPU simulation was executed with 60 cores and 10 GPU devices.}
\label{TOPHAT_DENS}
\end{figure*}

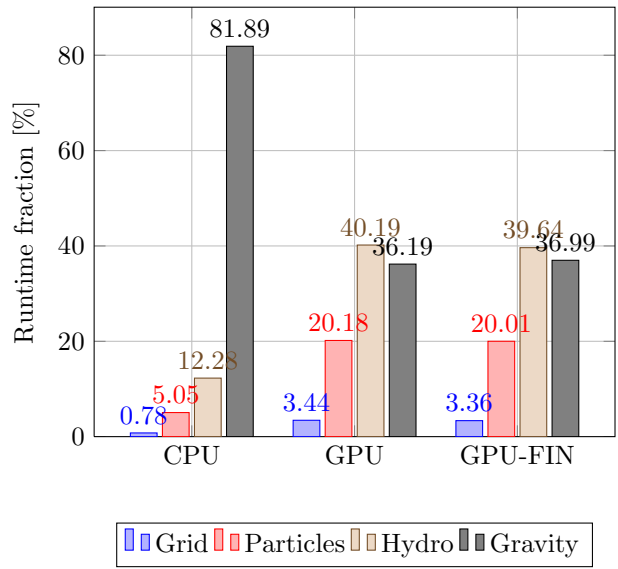
\begin{figure}[]
\begin{tikzpicture}
	\begin{axis}[
          xtick align=inside,
          ylabel={Runtime fraction [\%]},
          grid=major,
	ybar,
	enlarge x limits=0.3,
	bar width=10pt,
	nodes near coords,
         nodes near coords align={vertical},
	ymin=0,
	symbolic x coords={CPU, GPU, GPU-FIN},
	xtick=data,
	legend style={at={(0.5,-0.2)},
         anchor=north,legend columns=-1},
         xtick align=inside,
		]
		
   \addplot
coordinates
                  {   
(CPU,0.780)
(GPU, 3.437)	
(GPU-FIN,3.357)
};

\addplot
coordinates
		{
		(CPU, 5.052)
                   (GPU, 20.177)
                  (GPU-FIN,20.009)	
                 		};

	\addplot
coordinates
		{ 
		 (CPU,12.280)
                    (GPU, 40.193)	
                    (GPU-FIN,39.640)	
		};     \label{hy_bars}   		
				
\addplot
	         coordinates
		{
                (CPU, 81.887)
                (GPU, 36.193)	
                (GPU-FIN,36.992)	
		};  \label{grv_bars}  
	
         \legend{Grid, Particles, Hydro, Gravity}
	 \end{axis}
\end{tikzpicture}
\caption{Runtime fractions of different code parts for the CPU only (CPU) and GPU accelerated (GPU, GPU-FIN) SP\text{-}TL (CPU) simulations. The values for CPU and for GPU-FIN refer to the whole simulation time of the CPU only and the GPU accelerated simulation. The values for GPU refer to the GPU accelerated SP\text{-}TL simulation at a state near the CPU simulatio's final state. For the CPU only simulation, the gravity solver dominated the runtime with more than $80\%$, while for the GPU accelerated simulations, the hydro solver is the dominating module. (The GPU accelerated SP\text{-}TL simulation was run with 10 GPU devices and 60 CPU cores.)}
\label{PROP_LONG}
\end{figure}

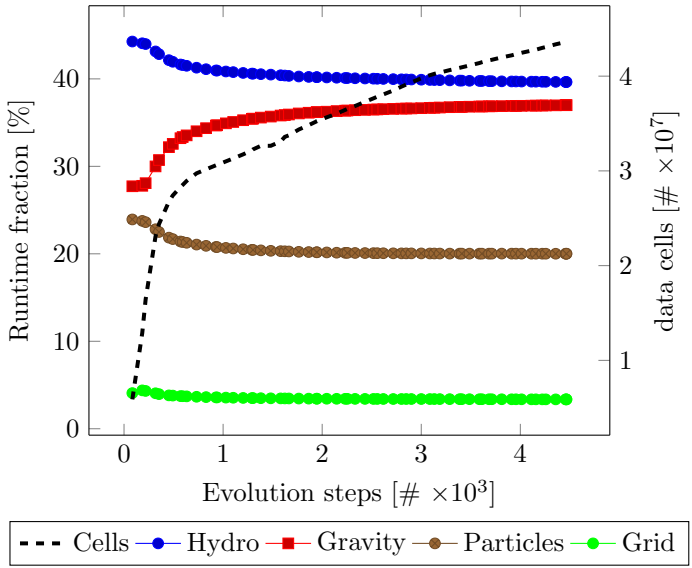
\begin{figure}[]
\begin{tikzpicture}
\pgfkeys{/pgf/number format/.cd,
sci,
sci generic={mantissa sep=\times,exponent={10^{#1}}}}
\begin{axis}[
          axis y line*=left,
          xlabel={Evolution steps [\# $\times 10^3$]},
          ylabel={Runtime fraction [\%]},
         yticklabel style={/pgf/number format/.cd, fixed},
         xtick={0,1000,2000,3000,4000}, 
         xticklabels={0,1,2,3,4}, 
]
\addplot table [x=STEP, y=HYDRO, col sep=comma] {TURBULENCELONGGPUEVO.csv};
\label{hy_long}
\addplot table [x=STEP, y=GRAVITY, col sep=comma] {TURBULENCELONGGPUEVO.csv};
\label{grv_long}
\addplot table [x=STEP, y=PARTICLES, col sep=comma] {TURBULENCELONGGPUEVO.csv};
\label{part_long}
\addplot [color=green,mark=*]table [x=STEP, y=GRID, col sep=comma] {TURBULENCELONGGPUEVO.csv};
\label{grid_long}
\end{axis}

\begin{axis}[
   axis y line*=right,
   ylabel={data cells [\# $\times 10^7$]},
  axis x line=none,
  yticklabel style={/pgf/number format/.cd, fixed},
  legend style={at={(0.5,-0.2)},
  anchor=north,legend columns=-1}
]
 \addplot[color=black,dashed,line width=1.5pt]
                table [x=STEP, y=CELL_NORM, col sep=comma] {TURBULENCELONGGPUEVO.csv};
                \label{cell_long}
                \addlegendentry{Cells}
                \addlegendimage{/pgfplots/refstyle=hy_long}\addlegendentry{Hydro}
                \addlegendimage{/pgfplots/refstyle=grv_long}\addlegendentry{Gravity}
                \addlegendimage{/pgfplots/refstyle=part_long}\addlegendentry{Particles}
                \addlegendimage{/pgfplots/refstyle=grid_long}\addlegendentry{Grid}
 \end{axis}

\end{tikzpicture}
\caption{Runtime fractions of the major modules used in the GPU accelerated SP\text{-}TL simulation. The fractional time of the accelerated gravity solver (\ref{grv_long}) grows as a result of the increasing number of data cells (\ref{cell_long}).}
\label{contrib_long_routines}
\end{figure}

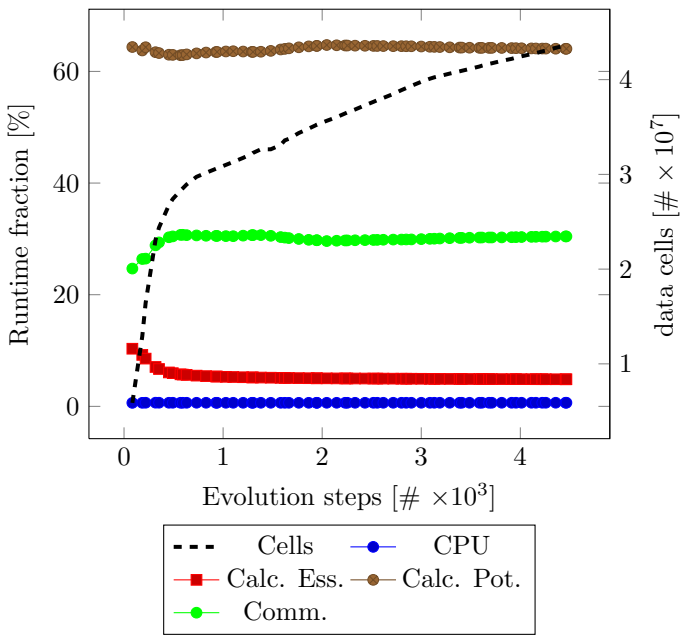
\begin{figure}[]
\begin{tikzpicture}
\begin{axis}[
          ylabel={Runtime fraction [\%]},
         xlabel={Evolution steps [\# $\times 10^3$]},
         legend style={at={(0.5,-0.2)},
         anchor=north,legend columns=-1},
        xtick={0,1000,2000,3000,4000}, 
         xticklabels={0,1,2,3,4}, 
]
\addplot table [x= STEP, y=CPU_PROP, col sep=comma] {TURBULENCELONGEVOROUTINES.csv};
\label{cpu_frac}
\addplot table [x=STEP, y=CALC_EXP_PROP, col sep=comma] {TURBULENCELONGEVOROUTINES.csv};
\label{exp_frac}
\addplot table [x=STEP, y=CALC_POT_PROP, col sep=comma] {TURBULENCELONGEVOROUTINES.csv};
\label{pot_frac}
\addplot [color=green,mark=*]table [x=STEP, y=WAIT_PROP, col sep=comma] {TURBULENCELONGEVOROUTINES.csv};
\label{wait_frac}

\end{axis}
\begin{axis}[
   axis y line*=right,
   ylabel={data cells [$\# \times 10^7$]},
  axis x line=none,
  yticklabel style={/pgf/number format/.cd},
  legend style={at={(0.5,-0.2)},
  anchor=north,legend columns=2}
]
 \addplot[color=black,dashed,line width=1.5pt]
                table [x=STEP, y=CELL_NORM, col sep=comma] {TURBULENCELONGGPUEVO.csv};
                \label{cell_norm}
                \addlegendentry{Cells}
                \addlegendimage{/pgfplots/refstyle=cpu_frac}\addlegendentry{CPU}
                \addlegendimage{/pgfplots/refstyle=exp_frac}\addlegendentry{Calc. Ess.}
                \addlegendimage{/pgfplots/refstyle=pot_frac}\addlegendentry{Calc. Pot.}
                \addlegendimage{/pgfplots/refstyle=wait_frac}\addlegendentry{Comm.}
 \end{axis}

\end{tikzpicture}
\caption{Average runtime fractions of selected routines of the GPU-BH tree solver during the evolution of the SP\text{-}TL simulation. The routines solely executed on the CPU (gathering data, allocation, deallocation, ...) stay at a constant fraction (\ref{cpu_frac}), while the fraction of the non overlapped communication time (\ref{wait_frac}) is highly affected by the increasing cell count (\ref{cell_norm}).}
\label{contib_comm}
\end{figure}

With the GPU accelerated simulation, $4465$ evolution steps were
executed and a simulation time of $t_\mathrm{s} \approx 1.6 \times
10^4\ \mathrm{yrs}$ was reached with a maximum density of $3.73 \times
10^{16}\ \mathrm{g}\ \mathrm{cm}^{-3}$. Similar to the CPU only
simulation, the maximum refinement level was not reached and no sink
particles were formed. Note, that the GPU-BH tree simulation reached
the $2006$ step mark already after $\approx 2.6\ \mathrm{days}$. An
overview of the simulation parameters for the final stages is given in
table~\ref{tab_finPAR}. For comparison, we show slices of the density distributions of
both simulations at the same simulation time ($t_\mathrm{s} = 10.2\,
\mathrm{kyr}$) in fig.~\ref{TOPHAT_DENS}.
\begin{table}[htb]%
\caption{Final simulation parameters}\renewcommand{\tabcolsep}{0.5pc}
\renewcommand{%
\arraystretch}{1.2}%
\small
\begin{tabular}{@{}llc}
\hline
\textbf{Parameter } & CPU-BH & GPU-BH \\ \hline
Max. refinement & 9 &10 \\
Leaf-blocks total & 69273 &    85261\\
Leaf-blocks/CPU & $\approx 962$ &   $\approx 1421$\\
Max. gas density & $9.12\times 10^{17}\ \mathrm{g\ cm}^{-3}$&$3.73 \times 10^{16}\ \mathrm{g\ cm}^{-3}$\\
Evolution steps &2006 &4465 \\
Max. sim. time&$3.2185\times 10^{11}\ \mathrm{s}$& $5.2345\times 10^{11}\ \mathrm{s}$ \\
\hline
\end{tabular}
\\[2pt]
Simulation parameters in the final stage (after 168 compute hours,
i.e. 7 days) of the GPU accelerated simulation and the CPU only simulation.
\label{tab_finPAR}
\end{table}
A simple comparison between the number of executed evolution steps
suggests a speedup factor for the GPU accelerated simulation of
$\approx 2.2$. At $t_s = 10.2\ \mathrm{kyr}$ where both simulations
hold approximately equal numbers of leaf blocks, we find a total
simulation speedup factor for the GPU-accelerated simulation of
$\approx 2.7$.\footnote{These estimations do not take into account
  the workload difference caused by different CPU counts.} Note that
this simulations ran with a $1/6$ GPU/CPU ratio, nevertheless, we
expect a much larger speedup for larger GPU/CPU ratios. 
Furthermore, in fig.~\ref{PROP_LONG} we show the runtime fractions of different
modules for both SP\text{-}TL simulations.  Following our
expectations from the previous simulations (\ref{hy_perc}), we find
the runtime fractions of our gravity solver and the hydro solver to be
roughly the same with $\approx 4\%$ difference (fig. ~\ref{PROP_LONG},
GPU(\ref{hy_bars},\ref{grv_bars}),
GPU-FIN(\ref{hy_bars},\ref{grv_bars}) ). With the CPU only
SP\text{-}TL simulation, the CPU-BH gravity solver is by far the
dominating module with over $80\%$ wall-time contribution.
In the GPU accelerated SP\text{-}TL simulation, the number of AMR
blocks increases with the simulation time and  as a result, the
workload for our GPU-BH tree solver grows. The runtime fractions of
the major routines executed during the evolution of the accelerated
SP\text{-}TL simulation are shown in
fig.~\ref{contrib_long_routines}. We see a growing runtime
contribution of our gravity solver during the first 1000 steps of
$\approx 8 \%$ from the initial $27 \%$ to $35 \%$. After 1000 steps,
the runtime contribution slowly converges to $\approx 36
\%$. Furthermore, we show the fractional contributions of the tree-algorithm
in fig.~\ref{contib_comm}. Here we see that the fractional increase
of the gravity solver (see fig.~\ref{contrib_long_routines}) comes
mainly from the additional communication time with the growing number
of data cells. The communication dominates the runtime during the
early steps ($N < 500$) due to small number of grid cells. 

\subsubsection{Strong scaling with the turbulent top-hat sphere}\label{Strong scaling with the turbulent top-hat sphere section}

\begin{figure}[]
\centering
\includegraphics[width=0.5\textwidth]{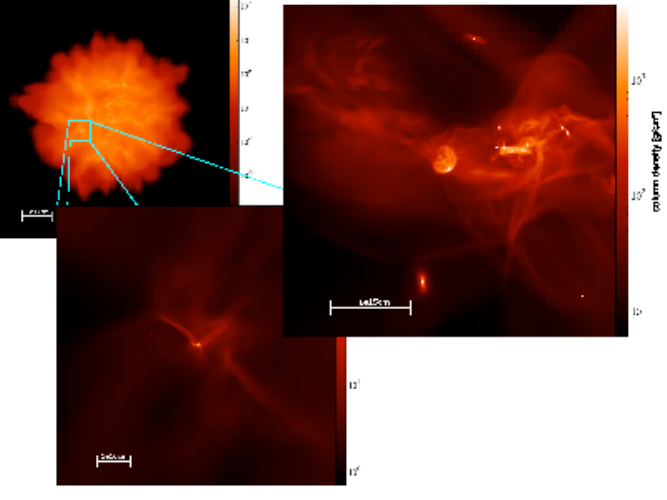}
\caption{The SP\text{-}TL simulation at $t_s=1.2 \times 10^{12}\ \mathrm{s}\  \approx 3.8 \times 10^4\ \mathrm{yrs}$. Plots show the column density $\mathrm{g\ cm}^{-2}$ for the lower collapse region at different resolutions. From left to right, the box length is reduced from $8.0 \times 10^{17}\ \mathrm{cm}$ over $8.0 \times 10^{16}\ \mathrm{cm}$ to $3.0 \times 10^{15}\ \mathrm{cm}$. Sink particles appear as white dots in the large image to the right.}
\label{turb_zoom1}
\end{figure}

\begin{figure}[]
\centering
\includegraphics[width=0.5\textwidth]{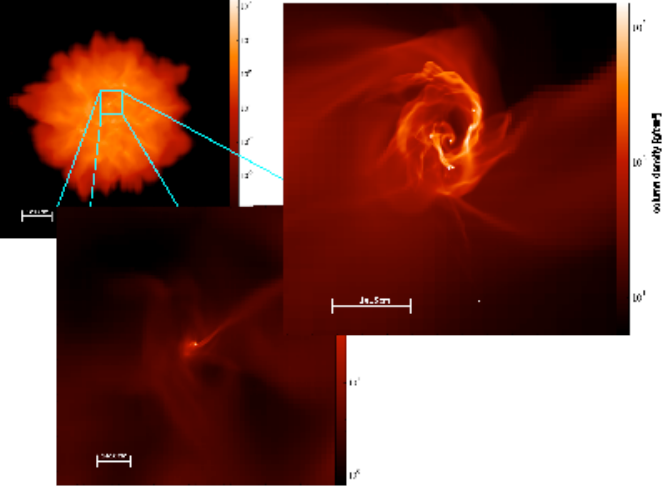}
\caption{The SP\text{-}TL simulation at $t_s=1.2 \times 10^{12}\ \mathrm{s} \approx 3.8 \times 10^4\ \mathrm{yrs}$. Plots show the column density $\mathrm{g\ cm}^{-2}$ for the upper collapse region at different resolutions. From left to right, the box length is reduced from $8.0 \times 10^{17}\ \mathrm{cm}$ over $8.0 \times 10^{16}\ \mathrm{cm}$ to $3.0 \times 10^{15}\ \mathrm{cm}$. Sink particles appear as white dots in the large image to the right.}
\label{turb_zoom2}
\end{figure}

We evaluate the strong scaling capabilities of our GPU-BH solver with
the turbulent top-hat sphere setup. For this, we run the
SP-TL simulation until
$t_ \mathrm{s} = 1.2 \times 10^{12}\ \sec \approx 3.8 \times 10^4\ \yr$. At his
point, two collapse regions and 26 sink particles with a total mass
$\approx 0.42\, \Msol$ are formed. We used a resolution of 12
data cells per Jeans length to scale the simulation to hold
$\approx 1.85 \times 10^7$ cells.  Figure~\ref{turb_zoom1} and
~\ref{turb_zoom2} show the two collapse regions at different
resolutions.

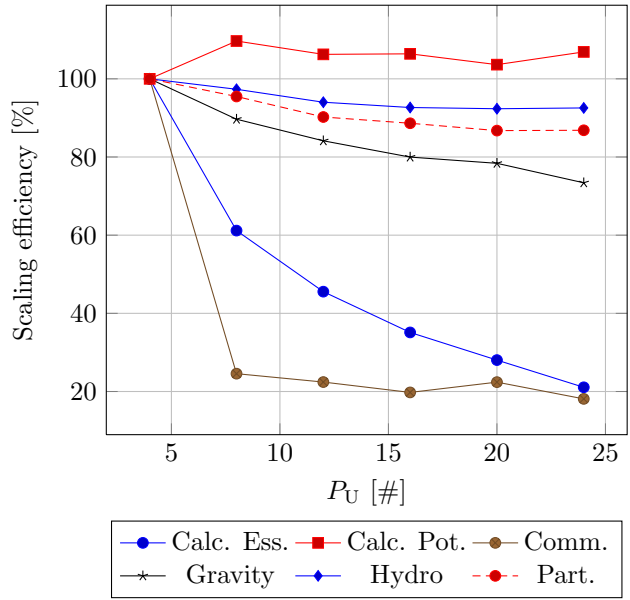
\begin{figure}[]
\begin{tikzpicture}
\begin{axis}[
           legend entries={Calc. Ess.,Calc. Pot. ,Comm. ,Gravity, Hydro, Part., Evo.},
          ylabel={Scaling efficiency [\%]},
          xlabel={$P_\mathrm{U}$ [\#]},
         legend style={at={(0.5,-0.2)},
         anchor=north,legend columns=3},
          grid=major,
]
\addplot table [x= NODES, y=EXP_SPD, col sep=comma] {TURBULENCELONGSTRONGPLT.csv};\label{ess_plt}
\addplot table [x= NODES, y=CALC_SPD, col sep=comma] {TURBULENCELONGSTRONGPLT.csv};
\addplot table [x= NODES, y=COM_SPD, col sep=comma] {TURBULENCELONGSTRONGPLT.csv};\label{com_plt}
\addplot table [x= NODES, y=GRAV_SPD, col sep=comma] {TURBULENCELONGSTRONGPLT.csv};\label{grv_plt}
\addplot table [x= NODES, y=HYDRO, col sep=comma] {TURBULENCELONGSTRONGPLT.csv};\label{hy_plt}
\addplot table [x= NODES, y=PART, col sep=comma] {TURBULENCELONGSTRONGPLT.csv};
\end{axis}

\end{tikzpicture}
\caption{Strong scaling efficiency in \% of linear for different code parts of the GPU-BH tree solver and the major evolution routines executed during the SP\text{-}TS simulations. One $P_\mathrm{U}$ refers to 6 CPU cores and 1 GPU. The strong scaling efficiency of our gravity solver (\ref{grv_plt}) drops down to $\approx 73\%$ of the linear scaling.}
\label{strong_turb}
\end{figure}

\begin{figure}[]
\begin{tikzpicture}
\begin{axis}[
           legend entries={CPU-BH, Gridsolver},
        ylabel={Speedup factor (Gravity)},
          xlabel={$P_\mathrm{U}$ [\#]},
         legend style={at={(0.5,-0.2)},
         anchor=north,legend columns=2},
          grid=major,
         nodes near coords,
]
\addplot table [x= NODES, y=SPD_WU, col sep=comma] {TURBULENCELONGGRVSPEEDUP.csv};\label{wu_plt}
\addplot table [x= NODES, y=SPD_GRD, col sep=comma] {TURBULENCELONGGRVSPEEDUP.csv};\label{grd_plt}

\end{axis}

\end{tikzpicture}
\caption{Strong scaling speedup factors of our accelerated gravity solver compared to the CPU-BH tree solver and the Gridsolver. The values are evaluated for the SP\text{-}TS simulation for the gravity unit only. One $P_\mathrm{U}$ refers to a unit of 6 CPU cores and 1 GPU. }
\label{strong_speed}
\end{figure}
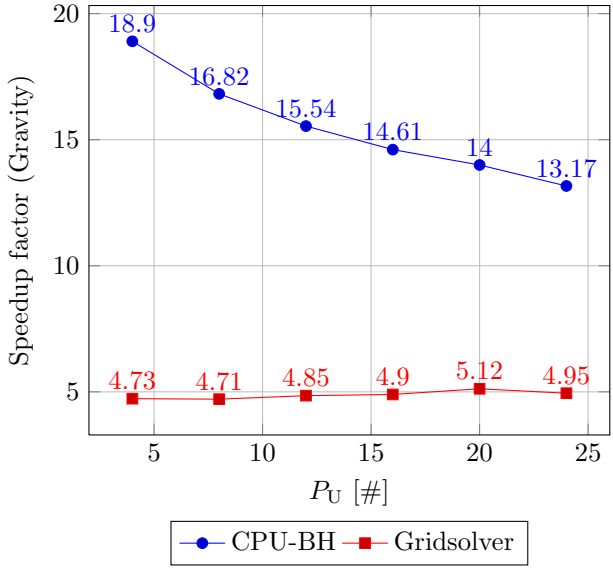

\noindent
We use the final checkpoint file of this simulation as a starting point for the strong scaling tests (SP\text{-}TS).\\
For the strong scaling, the problem size stays constant and we refer to the strong scaling efficiency (as a percentage of linear) with 
\begin{equation}\label{strong_eq}
\frac{t_1}{(N*t_N)}* 100\%,
\end{equation} 
where $t_1$ is the time spent in the gravity unit with one processing element, $N$ is the number of processing elements and $t_N$ is the time spent in the gravity unit with $N$ processing elements. Again, we use six CPU cores and one GPU device as a single processing unit ($P_\mathrm{U}$).\\
Figure~\ref{strong_turb} shows plots of the strong scaling efficiency of different code parts of our gravity solver and the main modules run during the simulation. The values are evaluated for 4, 8, 12, 16 , 20 and 24 $P_\mathrm{U}$. For each test, we measured the runtime for 2 evolution steps starting from the afore mentioned checkpoint file. The total scaling efficiency of our gravity solver dropped down to $\approx 73 \%$ for 24 $P_\mathrm{U}$. The loss in efficiency can be attributed to the routines for calculating the Essential Nodes (fig.~\ref{strong_turb},~\ref{ess_plt}), since these show the largest drop in efficiency. We trace this back to the increased loop size and increased memory operation cost as a result of the higher number of cores.\\
For comparison, we run the same simulation setup with the CPU-BH tree solver and the Gridsolver. Figure~\ref{strong_turb_solvers} shows the evaluated strong scaling efficiency for the respective simulations. We see a super linear scaling efficiency of the CPU-BH tree code (fig.~\ref{strong_turb_solvers},~\ref{wu_grv_strong_plt}) and sub linear strong scaling of our accelerated solver (fig.~\ref{strong_turb_solvers},~\ref{gpu_grv_strong_plt}) and the Gridsolver (fig.~\ref{strong_turb_solvers},~\ref{grid_grv_strong_plt}). The Gridsolver and our accelerated solver show an almost identical scaling behavior with a slight advantage of $\approx 3\ \%$ towards our solver.\newline
\noindent
Although the strong scaling efficiency of our solver shows a non linear behavior, we find a reasonable performance gain compared to the other solvers (see fig.~\ref{strong_speed}).
In comparison to the CPU-BH tree code, we find a high speedup of factor $19$ with 4 $P_\mathrm{U}$ and a speedup factor of $13$ with $24$ $P_\mathrm{U}$ (fig.~\ref{strong_speed},~\ref{wu_plt}). In comparison to the Gridsolver, our GPU accelerated solver was roughly 5 times faster for all numbers of $P_\mathrm{U}$ (fig.~\ref{strong_speed},~\ref{grd_plt}).\\
\begin{figure}[]
\begin{tikzpicture}
\begin{axis}[
           legend entries={GPU-BH,  CPU-BH,  GRID},
           ylabel={Scaling efficiency [\%] (Gravity)},
          xlabel={$P_\mathrm{U}$ [\#]},
          legend style={at={(0.5,-0.2)},
          anchor=north,legend columns=3},
          grid=major,
]
\addplot table [x= NODES, y=GPU_GRV_STRONG, col sep=comma] {TURBULENCELONGEVOSOLVERSSTRONG.csv};
\label{gpu_grv_strong_plt}

\addplot table [x= NODES, y=WU_GRV_STRONG, col sep=comma] {TURBULENCELONGEVOSOLVERSSTRONG.csv};
\label{wu_grv_strong_plt}

\addplot table [x= NODES, y=GRID_GRV_STRONG, col sep=comma] {TURBULENCELONGEVOSOLVERSSTRONG.csv};
\label{grid_grv_strong_plt}

\end{axis}

\end{tikzpicture}
\caption{Strong scaling efficiency for the gravity unit in \% of linear with the SP\text{-}TS setup for our GPU accelerated gravity solver (\ref{gpu_grv_strong_plt}) the CPU-BH tree solver (\ref{wu_grv_strong_plt}), and the Gridsolver (\ref{grid_grv_strong_plt}). The CPU-BH tree solver shows a super linear strong scaling behavior (\ref{wu_grv_strong_plt}), while our accelerated solver (\ref{gpu_grv_strong_plt}) and the Gridsolver (\ref{grid_grv_strong_plt}) show non ideal scaling capabilities. One $P_\mathrm{U}$ refers to 6 CPU cores and 1 GPU.}
\label{strong_turb_solvers}
\end{figure}
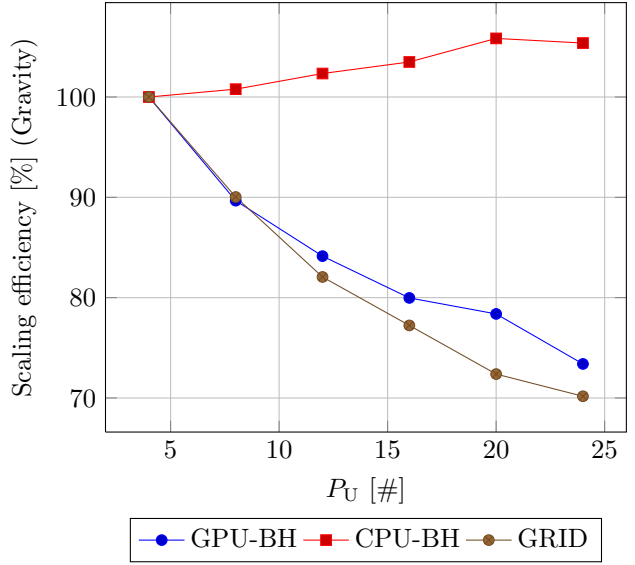

\begin{figure}[]
\begin{tikzpicture}
\begin{axis}[
           legend entries={GPU-BH,  CPU-BH,  GRID},
           ylabel={Data throughput [$10^5 \times \mathrm{cells} / \mathrm{sec}$]},
          xlabel={$P_\mathrm{U}$ [\#]},
          legend style={at={(0.5,-0.2)},
          anchor=north,legend columns=3},
          grid=major,
]
\addplot table [x= NODES, y=CELLS_PSEC, col sep=comma] {TURBULENCELONGSTRONGSCALINGROUTINES.csv};
\label{gpu_strong_work}
\addplot table [x= NODES, y=CPS, col sep=comma] {TURBULENCELONGWUNSCHEVO.csv};
\label{wu_strong_work}
\addplot table [x= NODES, y=CPS, col sep=comma] {TURBULENCELONGGRIDEVO.csv};
\label{grid_strong_work}
\end{axis}

\end{tikzpicture}
\caption{Strong scaling data throughput in data cells/sec of the different gravity solvers for the SP\text{-}TS simulations. With the CPU-BH tree solver (\ref{wu_strong_work}) and the Gridsolver (\ref{grid_strong_work}) , we reached the maximum data throughput of $5 \times 10^4$ respectively $1.16 \times 10^5$ data cells per second with 24 $P_\mathrm{U}$. A similar value of $\approx 1.97 \times 10^5$ data cells per second is reached with our accelerated gravity solver (\ref{gpu_strong_work}) with only 4 $P_\mathrm{U}$. For our accelerated solver, we reached the maximum of nearly $9 \times 10^5$ data cells per second with 24 $P_\mathrm{U}$. One $P_\mathrm{U}$ refers to 6 CPU cores and 1 GPU.}
\label{strong_turb_data}
\end{figure}
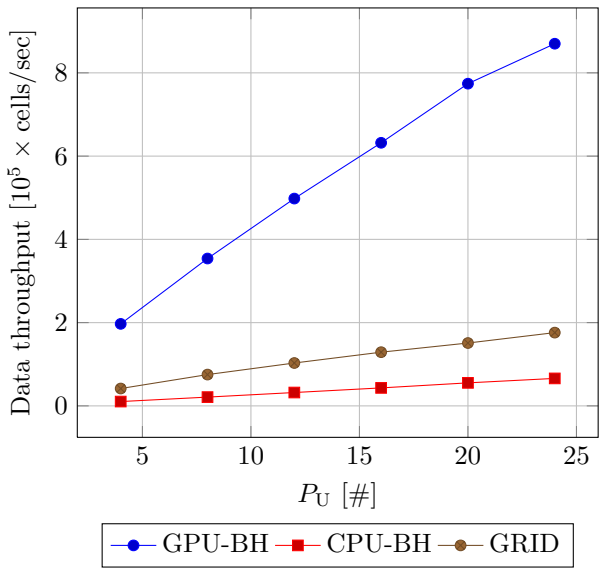

\begin{figure}[]
\begin{tikzpicture}
\begin{axis}[
           legend entries={CPU-BH 24 Cores, CPU-BH 144 Cores, GRID 24 Cores, GRID 144 Cores},
          xticklabels={ ,12,6,4,3, ,2},
          xtick={0,2,4,6,8,10,12},
          xlabel={GPU [\#]},
         ylabel={Speedup factor (Gravity)},
         legend style={at={(0.5,-0.2)},
         anchor=north,legend columns=2},
          grid=major,
          x dir=reverse,
]
\addplot table [x= NODES, y=UP_WU_24C, col sep=comma] {TURBULENCELONGSPEEDRATIO.csv};\label{wu_plt_24}
\addplot table [x= NODES, y=UP_WU_144C, col sep=comma] {TURBULENCELONGSPEEDRATIO.csv};\label{wu_plt_144}
\addplot table [x= NODES, y=UP_GRID_24C, col sep=comma] {TURBULENCELONGSPEEDRATIO.csv};\label{grid_plt_24}
\addplot table [x= NODES, y=UP_GRID_144C, col sep=comma] {TURBULENCELONGSPEEDRATIO.csv};\label{grid_plt_144}

\end{axis}

\end{tikzpicture}
\caption{Speedup factors of our accelerated gravity solver run on 24 cores with different CPU/GPU ratios compared to the CPU-BH tree solver and the Gridsolver run on 24 and 144 CPU cores. The values are evaluated fort he SP\text{-}TS simulation. The X-axis refers to the GPU count for different GPU/CPU ratios using a total of 24 CPU cores.}
\label{turb_ratio}
\end{figure}
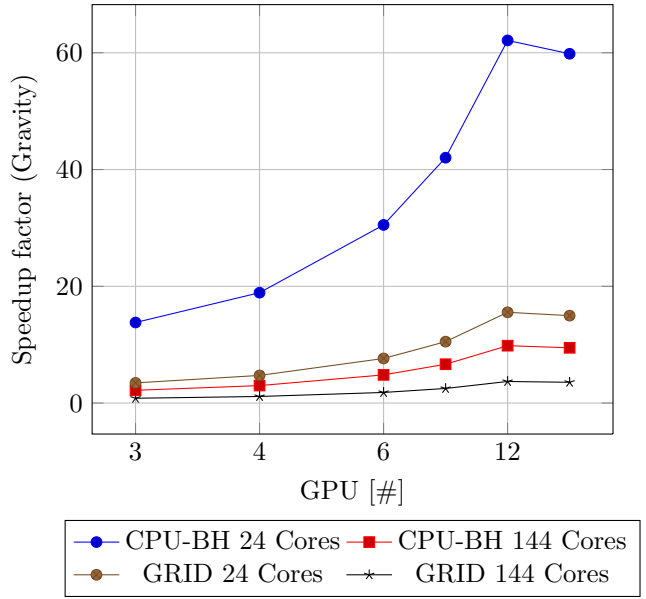

\begin{figure}
\begin{tikzpicture}
 \begin{axis}[
symbolic x coords={3,4,6,8,12,24},
    xlabel={GPU [\#]},
    ylabel={Speedup factor (Evolution)},
    scaled x ticks=false,
   nodes near coords,
    legend pos=north west,
    legend entries={CPU-BH ,Gridsolver},
    legend style={at={(0.5,-0.2)},
    anchor=north,legend columns=2}
]
\addplot [
color=red, 
mark = *] 
table [
x = CPU,
y = SPUP
   ] {

     SPUP CPU
      10.510  24
      9.874    12
      9.009      8  
      8.179      6 
      5.948      4 
      5.616      3
};
\label{Wu24_plot}
\addplot [
color=blue,
mark = square*
]
table[
x = CPU,
y = SPUP 
] {

     SPUP CPU
      3.025    24
      2.842    12
      2.593      8  
      2.354      6 
     1.712       4 
     1.616       3
};
\label{grid24_plot}

\end{axis}
\end{tikzpicture}
\caption{Evolution step speedup factors of our accelerated gravity solver with varying CPU/GPU ratios and a constant number of CPU cores compared to the CPU-BH code (\ref{Wu24_plot}) and the Gridsolver (\ref{grid24_plot}). The X-axis refers to the GPU count for different GPU/CPU ratios using a total of 24 CPU cores. All values are evaluated for the SP\text{-}TS simulation run with 24 CPU cores. }
\label{turb_total_ratio}
\end{figure}
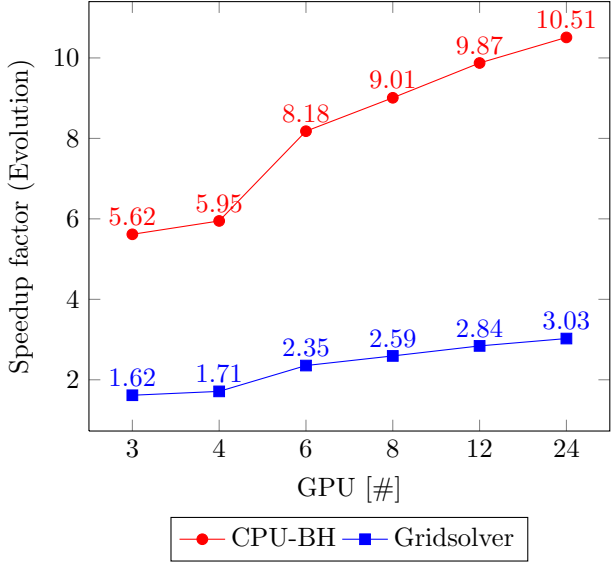

\noindent
Following the speedup factor, we see a much higher data throughput with our accelerated solver than with the other solvers (see fig.~\ref{strong_turb_data}). For our accelerated solver, we evaluated a data throughput of $\approx 2.0 \times 10^5$ data cells/sec with just 4 $P_\mathrm{U}$ (fig.~\ref{strong_turb_data},~\ref{gpu_strong_work}). At the same cores, we find much lower data throughput of $\approx 4.7 \times 10^4$ cells/sec for the Grdsolver and $\approx 1.0 \times 10^4$ cells/sec for the GPU-BH tree solver. We find the maximum data throughput values with $\approx 6.6 \times 10^4$ cells/sec with the CPU-BH solver $\approx 1.7 \times 10^5$ cells/sec with the Gridsolver and  $\approx 8.7 \times 10^5$ cells/sec with our accelerated GPU-BH solver for $24 P_\mathrm{U}$ . Note that even with additional 120 cores ($20 P_\mathrm{U}$) both the Gridsolver and the CPU-BH tree solver stayed below the $\approx 2.0 \times 10^5$ data cells/sec mark  which our solver reached with 4 $P_\mathrm{U}$.
The values reported in figure~\ref{strong_speed} only refer to a GPU/CPU ratio of 1/6 and we find higher speedup factors when utilizing less CPU core for each GPU. Figure~\ref{turb_ratio} shows the calculated speedup factors for different ratios. We find the highest speedup of factor $63$ for our GPU-BH tree solver compared to the CPU-BH tree solver when run with the same number of CPU cores and a CPU/GPU ratio of 2/1 (12 GPU devices and 14 CPU Cores). Even when utilizing 144 (6 times more) CPU cores to the CPU-BH solver, we find our GPU accelerated solver to be over $9$ times faster. 
We measure the speedup based on the ``evolution'' time stamp in the FLASH log file. Hence we ignore the initialization which is negligible in production simulations \citep[][]{Cordery:2014eq}. As can be seen in fig. \ref{turb_ratio}, we find the best speedup with a one-to-one CPU/GPU ratio (24 CPU Cores and 24 GPU devices). Here, we see a total speedup of factor 10 compared to the CPU-BH solver and a speedup of 3 compared to the gravity Gridsolver (see fig. \ref{turb_total_ratio}).

\section{Conclusions}
We described a GPU accelerated BH tree code as an additional Poisson solver for the FLASH4 software package. Our implementation works only
for three-dimensional problems. Furthermore, the tree walk is limited
to a maximum number of $\approx 32 \times 10^6$ data cells for each
process.\footnote{Assuming a GPU with at least 6GB of memory} Three
different simulation setups were used to test the accuracy and the
performance of our novel GPU-BH code. Depending on the setup and
GPU/CPU ratio we find a speedup for the gravity unit of at least a factor of 3 and up to 60
compared to the original CPU-BH implementation. For GPU/CPU ratios below 1/6 
the runtime of our simulations is no more dominated by the gravity unit but by the hydro solver. 
Hence we find lower speedup factors between 1.6 and 10  for the total application runtime. 
We have shown that
even with a small GPU/CPU ratio an advantageous performance gain can
be achieved.
The GPU-BH code was written for GPU-devices with a minimum
compute capability of 2.0 (Fermi architecture) and we expect further
runtime improvements by porting the code onto more modern devices with a higher compute
capability. Here, we expect improvements not only from higher clock rates and higher register counts, but from additional features. E.g. with a compute capability $\ge 3.0$ the warp shuffle functions {\tt{shfl()}} could be used to reduce the shared memory usage in the force calculation kernels.\\
Since our GPU code is written in CUDA, only NVIDIA GPUs are
supported with the current version. A portable version of the GPU-BH
tree code ported to OpenCL will be available in a later release. The
GPU-BH tree solver for FLASH4 is available for free from the Hamburg
Observatory at www.hs.uni-hamburg.de/gpubh.
\section{Acknoledgement}
The PARAMESH software used in this work was developed at the NASA Goddard Space Flight Center and Drexel University under NASA's HPCC and ESTO/CT projects and under grant NNG04GP79G from the NASA /AISR project. The software used in this work was in part developed by the DOE NNSA-ASC OASCR Flash Center at the University of Chicago. This work was funded by the Deutsche Forschungsgemeinschaft under grant BA 3707/4-1. Much of the analysis and data visualization was performed using the yt toolkit by \citep{2011ApJS..192....9T}
\appendix

\section{Periodic boundaries}
The GPU-BH tree code supports isolated and periodic boundary conditions. In case of periodic boundary conditions, a node's or data cell's contribution to the gravitational potential of a data cell $d_c$ changes to
\begin{equation}
\phi=GM\ f_{\sm{EF}}(\vec r).
\end{equation}
Here, $G$ is the gravitational constant, $M$ is the mass of the node or data cell, $\vec r$ is the position vector between $d_c$ and data cell or node and $f_\mathrm{EF}$ is the Ewald field. The original Ewald method is a method for computing the gravitational field for problems with periodic boundary conditions in three directions. Using the original method, the gravitational potential is split into two parts,
\begin{equation}
G m/r = Gm\ \mathrm{erf}(\alpha r)/r + Gm\  \mathrm{erfc}(\alpha r)/r
\end{equation} 
where $\alpha$ is an arbitrary constant. By applying Poisson summation formula on the erfc terms, the gravitational field at position $\vec r$ can be written in the form
￼￼￼￼￼￼￼￼￼￼￼\begin{equation}\label{ewald}
\begin{split}
\Phi(\vec r)=&-G\sum _{a=1}^N m_\mathrm{a} \times \\
& \left( \sum _\mathrm{i_1, i_2, i_3} A_\mathrm{S}(\vec r, \vec r_\mathrm{a}, \vec l_\mathrm{i_1,i_2,i_3}) + A_L(\vec r, \vec r_\mathrm{a}, \vec l_\mathrm{i_1,i_2,i_3}) \right).\\
\end{split}
\end{equation} 
The first sum runs over whole computational domain, where mass $m_\mathrm{a}$  is at position $\vec r_\mathrm{a}$. The second  sum runs over all neighboring computational domains, which are at positions $\vec l_\mathrm{i_1,i_2,i_3}$ and $ A_\mathrm{S}(\vec r, \vec r_\mathrm{a}, \vec l_\mathrm{i_1,i_2,i_3})$ and $A_\mathrm{L}(\vec r, \vec r_\mathrm{a}, \vec l_\mathrm{i_1,i_2,i_3})$ are short, resp. long-range contributions. The Ewald field is calculated once on startup and stored in a large array representing a hierarchy of nested grids. The evaluation of the Ewald Field at certain points is only needed during the final force evaluation and carried out during the respective tree walk using quadratic interpolation.  

\section{Usage}
The GPU-BH tree code for FLASH4 implements a set of runtime parameters to control the accuracy and runtime. All the parameters can be set individually using the common $flash.par$ file. The opening angle parameter $\theta$ can be set  with \emph{grv\_bh\_gpuLimAngle}
to a value between $0.1$ and $1.0$. The gathered data strategy can be selected with setting the  runtime parameter \emph{grv\_bh\_gpu\_concat} to $1$ where the default value of $0$ refers to a serialized GPU device access pattern. In case periodic boundaries are used, the respective Ewald Field is controlled via the runtime parameters \emph{gpuEwaldSeriesN} which controls the range of the indices $\mathrm{i_1,i_2,i_3}$ in (Eq.~\ref{ewald}) and \emph{gpuEwaldSeriesNref} which refers to the number of nested grids in the field array.
The number of data points in each grid can be set with the parameters \emph{gpuEwaldFieldNx}, \emph{gpuEwaldFieldNy} and  \emph{gpuEwaldFieldNz}. Note, that the Ewald field resides in GPU memory during the hole simulation.



\bibliographystyle{plainnat}
\bibliography{bib.bib}







\end{document}